\documentclass[aps, prd, showpacs, superscriptaddress, nofootinbib, onecolumn]{revtex4}
\usepackage{amssymb,amsmath,microtype}
\usepackage{graphicx}

\newtheorem{thm}{Conjecture}

\def\be{\begin{equation}}
\def\ee{\end{equation}}
\def\bea{\begin{eqnarray}}
\def\eea{\end{eqnarray}}

\begin{document}

\title{Evolution of cosmological perturbations and the production of non-Gaussianities through a nonsingular bounce:
Indications for a no-go theorem in single field matter bounce cosmologies}

\author{Jerome Quintin}
\email{jquintin@physics.mcgill.ca}
\affiliation{Department of Physics, McGill University, Montr\'eal, Qu\'ebec H3A 2T8, Canada}

\author{Zeinab Sherkatghanad}
\email{z.sherkat@mail.mcgill.ca}
\affiliation{Department of Physics, McGill University, Montr\'eal, Qu\'ebec H3A 2T8, Canada}
\affiliation{Department of Physics, Isfahan University of Technology, Isfahan, 84156-83111, Iran}

\author{Yi-Fu Cai}
\email{yifucai@ustc.edu.cn}
\affiliation{CAS Key Laboratory for Researches in Galaxies and Cosmology, Department of Astronomy, University of Science and Technology of China, Chinese Academy of Sciences, Hefei, Anhui 230026, China}

\author{Robert H. Brandenberger}
\email{rhb@hep.physics.mcgill.ca}
\affiliation{Department of Physics, McGill University, Montr\'eal, Qu\'ebec H3A 2T8, Canada}

\begin{abstract}

Assuming that curvature perturbations and gravitational waves originally
arise from vacuum fluctuations in a matter-dominated phase of contraction,
we study the dynamics of the cosmological perturbations evolving through a
nonsingular bouncing phase described by a generic single scalar field
Lagrangian minimally coupled to Einstein gravity. In order for such a model
to be consistent with the current upper limits on the tensor-to-scalar ratio,
there must be an enhancement of the curvature fluctuations during the bounce
phase. We show that, while it remains possible to enlarge the amplitude of
curvature perturbations due to the nontrivial background evolution, this
growth is very limited because of the conservation of curvature perturbations
on super-Hubble scales. We further perform a general analysis of the
evolution of primordial non-Gaussianities through the bounce phase. By
studying the general form of the bispectrum we show that the non-Gaussianity
parameter $f_{\mathrm{NL}}$ (which is of order unity before the bounce phase)
is enhanced during the bounce phase if the curvature fluctuations grow. Hence,
in such nonsingular bounce models with matter given by a single scalar field,
there appears to be a tension between obtaining a small enough
tensor-to-scalar ratio and not obtaining a value of $f_{\mathrm{NL}}$ in
excess of the current upper bounds. 
{\it{This conclusion may be considered as a ``no-go'' theorem
that rules out any single field matter bounce cosmology starting with
vacuum initial conditions for the fluctuations.}}

\end{abstract}


\maketitle
\tableofcontents

\section{Introduction}\label{sec:intro}

As was realized in \cite{Wands:1998yp,Finelli:2001sr}, there is a duality
between the evolution of curvature fluctuations in an exponentially
expanding universe and in a contracting universe with the
equation of state of matter. In both cases, curvature fluctuations which originate as
quantum vacuum perturbations on sub-Hubble scales acquire a scale-invariant
spectrum at later times on super-Hubble scales. The observed small red tilt
of the spectrum of curvature perturbations which has now been confirmed by
observations (see e.g.\ \cite{Ade:2013zuv,Ade:2015xua}) can be obtained in an
expanding universe by a slow decrease of the Hubble constant during the
period of inflation\ \cite{Mukhanov:1981xt}, whereas in a matter-dominated phase of
contraction a small cosmological constant (with magnitude comparable to what
is needed to explain today's dark energy) yields the same tilt
\cite{Cai:2014jla} (see alternatively\ \cite{deHaro:2015wda}).
To avoid reaching a singularity at the end of the
contracting phase, it is necessary to either modify gravity or consider
matter violating the null energy condition (NEC). Then it is possible to
obtain nonsingular bouncing cosmologies which have the potential to yield
an explanation for the structures in the universe which we now observe. This
scenario of structure formation alternative to inflation is called the
``matter bounce'' scenario (see e.g.\ \cite{Brandenberger:2012zb,Brandenberger:2010dk} for
reviews).

Examples of modified gravity models which yield bouncing cosmologies include
the ``nonsingular Universe'' construction of\ \cite{Mukhanov:1991zn,Brandenberger:1993ef}, nonlocal gravity
actions like\ \cite{Biswas:2005qr}, or Ho\v{r}ava-Lifshitz gravity\ \cite{Brandenberger:2009yt}. It is
in general very hard to study the evolution of fluctuations in these models.
We will hence focus on models in which the bounce is obtained from the matter
sector. One method of obtaining a nonsingular bounce with a single scalar
field involves the formation of a ghost condensate during the bounce phase
(see\ \cite{Buchbinder:2007ad,Creminelli:2007aq,Lin:2010pf,LevasseurPerreault:2011mw,Qiu:2011cy,Easson:2011zy}
for initial developments). A general problem for bouncing cosmologies is the
Belinsky-Khalatnikov-Lifshitz (BKL) instability\ \cite{Belinsky:1970ew}, the fact that the energy density in
the form of anisotropies will explode and destroy the homogeneous bounce
\cite{Cai:2013vm}. This problem can be ``solved'' by endowing the scalar
matter field with a negative potential which leads to an
Ekpyrotic phase of contraction before the bounce
\cite{Cai:2012va,Cai:2013kja,Quintin:2014oea} and hence can mitigate the
anisotropy problem\ \cite{Erickson:2003zm} \footnote{Such a negative potential
may arise from the standard model Higgs field since, based on the recent Higgs
and top quark mass measurements, the standard model Higgs develops an
instability at large field values (in the absence of new physics)
\cite{Brandenberger:2015nua}.}.

In the matter bounce scenario, primordial quantum fluctuations exit the
Hubble horizon while the universe is in a matter-dominated contracting phase
and the resulting power spectrum of curvature perturbations is
scale-invariant\ \cite{Wands:1998yp,Finelli:2001sr}. On the other hand, the
gravitational wave mode obeys the same equation of motion on super-Hubble
scales as the curvature perturbations (considering the canonical variables
in each case). Hence, before the bounce phase the tensor-to-scalar ratio
$r$ would be of order unity. Thus, if the perturbations passed through the
nonsingular bounce unchanged, it would imply that curvature perturbations
and primordial gravitational waves would have the same amplitude after the
bounce. In terms of the tensor-to-scalar ratio, it would mean that
$r\sim\mathcal{O}(1)$, well above current observational bounds
\cite{Ade:2013zuv,Ade:2015xua,Ade:2015tva}.

Since the curvature fluctuations couple nontrivially to matter during the
bounce phase, whereas the tensor perturbations are determined simply by
the evolution of the scale factor $a(t)$, one may expect that the curvature
perturbations would be enhanced relative to the tensor modes during the
bounce. In fact, early calculations indicated that curvature perturbations
grew exponentially during the bounce phase, hence suppressing the
tensor-to-scalar ratio\ \cite{Cai:2012va,Cai:2014xxa}. A more recent
study\ \cite{Battarra:2014tga} numerically explored the evolution of
scalar fluctuations through a nonsingular bounce model similar to the
one studied in\ \cite{Cai:2012va} and found no enhancement of
curvature perturbations through the bounce. In light of the relevance of a
possible enhancement of the curvature fluctuations for the predicted value
of the tensor-to-scalar ratio, the growth of curvature fluctuations during a
nonsingular bounce needs to be reconsidered. This is what we aim to do in
this paper.

The second goal of this paper is to carefully track the evolution of the
three-point function (bispectrum) of curvature perturbations through the
bounce. In earlier work\ \cite{Cai:2009fn} it was shown that the bispectrum
of curvature fluctuations before the bounce phase has an amplitude of the
order $f_{\mathrm{NL}}\sim\mathcal{O}(1)$ with a specific shape. As we
argued above, if the perturbations were to pass through the nonsingular
bounce unchanged, it would imply a large tensor-to-scalar ratio in excess
of the observational bounds. On the other hand, if curvature perturbations
were to experience a nontrivial growth through the bounce, one should
expect additional nonzero contributions to the bispectrum coming from the
bounce phase, and there would then be the danger that the final amplitude
of the bispectrum exceeds the observational upper bounds from
\cite{Ade:2013ydc,Ade:2015ava}. Thus, a potential conflict looms: either
the tensor-to-scalar ratio is too large, or else the non-Gaussianities
exceed observational bounds. This problem has indeed already been found
in a model of a nonsingular bouncing cosmology in which a nonvanishing
positive spatial curvature is responsible for the bounce
\cite{Gao:2014hea,Gao:2014eaa}. We will study this issue in the context of
the more realistic models in which the nonsingular bounce is generated by
the matter sector. In particular, we will explore the question in the
context of a ghost-condensate bounce.

We will indeed demonstrate that, at least in our model, the evolution of
the curvature perturbations in the bounce phase connects the value of the
tensor-to-scalar ratio with the amplitude of non-Gaussianities. The
suppression of the tensor-to-scalar ratio to restore compatibility with the
observational bounds requires an enhancement of the curvature fluctuations
during the bounce phase. Such an enhancement will increase the magnitude of
the non-Gaussianities to a level inconsistent with the observational bounds
on the amplitude of the bispectrum. Based on our result we conjecture that
there exists a ``no-go'' theorem in single field nonsingular matter bounce
cosmologies which relates the tensor-to-scalar ratio and non-Gaussianities,
preventing these models to satisfy the current observational bounds. A
tensor-to-scalar ratio below current observational bounds would imply a too
large amplitude of non-Gaussianities, whereas non-Gaussianities of order
$f_{\mathrm{NL}}\sim\mathcal{O}(1)$ would imply a too large amplitude of
the primordial gravitational wave spectrum.
Therefore, a single field nonsingular matter bounce cannot be made
consistent with current observations if the primordial
perturbations arise from vacuum initial conditions.

Our analysis assumes that both curvature perturbations and gravitational
waves originate as quantum vacuum fluctuations in the initial phase of
contraction. A model with thermal fluctuations (as obtained for example in
the context of string gas cosmology\ \cite{Brandenberger:1988aj,Biswas:2006bs}) will easily avoid
our ``no-go'' theorem. As shown in\ \cite{Nayeri:2005ck,Brandenberger:2006vv,Brandenberger:2006xi,Brandenberger:2014faa}, we obtain a
tensor-to-scalar ratio much smaller than order unity while obtaining
non-Gaussianities which are negligible on cosmological scales\ \cite{Chen:2007js}.

The paper is organized as follows. We first start with a short review of
cosmological perturbation theory in Sec.\ \ref{sec:cosmopert}. We then
motivate the idea of the no-go theorem proposed in this paper in
Sec.\ \ref{sec:outlinenogo}. In Sec.\ \ref{sec:bounce}, we briefly review
the general picture of bouncing cosmology in terms of a single scalar field
of Galileon type. After that, in Sec.\ \ref{sec:pert} we analyze the
perturbation equation for primordial curvature perturbations at linear
order during the nonsingular bouncing phase. We point out under which
conditions there can be an enhancement of their amplitude. Then in
Sec.\ \ref{sec:nong}, we perform a detailed analysis of the bispectrum
generated in the bouncing phase of our specific model.
We combine the analyses of scalar and tensor perturbations
together with non-Gaussianities in Sec.\ \ref{sec:constraints},
and we show how current observational bounds severely constrain
the parameter space of the single field bouncing model.
The analysis is expected to hold quite
generally for single field matter bounce cosmologies. We conclude with a
discussion in Sec.\ \ref{sec:conclusion}. Throughout this paper, we adopt
the mostly minus convention for the metric and define the reduced Planck
mass as $M_p^2 \equiv 1/8\pi G_{\mathrm{N}}$ where $G_{\mathrm{N}}$ is
Newton's gravitational constant.

\section{A brief review of cosmological perturbation theory}\label{sec:cosmopert}

Linear perturbations of the metric about a homogeneous and isotropic
background space-time can be decomposed into scalar, vector, and tensor
modes (see\ \cite{Mukhanov:1990me} for a review of the theory of cosmological
perturbations and\ \cite{Brandenberger:2003vk} for an introductory overview). The
scalar modes are those which couple to matter energy density and pressure
perturbations. We call these the {\it cosmological perturbations}. Tensor
modes exist in the absence of matter - they correspond to gravitational
waves. In the case of matter without anisotropic stress at linear order in
the amplitude of the fluctuations, there is only one physical degree of
freedom for the scalar fluctuations. For the purpose of computations it is
often convenient to work in the conformal Newtonian gauge (coordinate system)
in which the perturbed metric for scalar modes reads
\begin{equation}
 ds^2=a^2(\eta)\left(\left[1+2\Phi(\eta,\vec{x})\right]d\eta^2-\left[1-2\Phi(\eta,\vec{x})\right]d\vec{x}^2\right)~,
\end{equation}
where $\eta$ denotes conformal time, $a(\eta)$ is the cosmological
scale factor, $\vec{x}$ represents comoving spatial coordinates,
and $\Phi$ denotes the gravitational potential.
For tensor modes, the perturbed metric reads
\begin{equation}
 ds^2=a^2(\eta)\left(d\eta^2-\left[\delta_{ij}+h_{ij}(\eta,\vec{x})\right]dx^idx^j\right)~,
\end{equation}
where $h_{ij}$ is trace-free and divergenceless.

Let us consider the matter content to be described by a single scalar field
of canonical form with Lagrangian density
\begin{equation} \label{canAction}
 \mathcal{L}_m = \frac{1}{2}M_p^2g^{\mu\nu}\nabla_{\mu}\phi\nabla_{\mu}\phi-V(\phi)~.
\end{equation}
Note that we take the scalar field to be dimensionless throughout this
paper as a convention. Linear perturbations of the scalar field then
have the form
\be
\phi(\eta,\vec{x}) = \phi_0(\eta)+\delta\phi(\eta,\vec{x}) \, ,
\ee
where $\phi_0$ is the unperturbed homogeneous part of $\phi$.
In the scalar sector, metric and matter perturbations couple to one another,
so it is useful to define a linear combination of these perturbations,
\begin{equation}
 \mathcal{R} \equiv \frac{\mathcal{H}}{\phi_0'}\delta\phi+\Phi~ \, .
\end{equation}
There are two reasons for focusing on this variable. First of all, it
gives the curvature fluctuation in comoving coordinates (coordinates
in which the matter field is uniform), and is hence the variable
we are interested in computing. Second, it is simply related
to the Sasaki-Mukhanov\ \cite{Sasaki:1986hm,Mukhanov:1988jd} variable $v$
in terms of which the action for cosmological
perturbations has canonical form. Note that in the above,
$\mathcal{H}\equiv a'/a$ is the conformal Hubble parameter
and a prime denotes a derivative with respect to conformal time.
In fact, the Sasaki-Mukhanov variable is
\be
v \equiv z\mathcal{R} \, ,
\ee
with
\begin{equation}
 z = a\frac{\phi_0'}{\mathcal{H}}M_p \, .
 \label{zcanonical}
\end{equation}

The equation of motion that results from expanding the perturbed action
for gravity and matter to second order is given by
\begin{equation}
\label{EOMvk}
 v_k''+\left(c_s^2k^2-\frac{z''}{z}\right)v_k = 0~.
\end{equation}
The equation is written in Fourier space, where $k$ represents the comoving
wave number of the curvature perturbations, and $c_s$ is the speed of sound
which is equal to one for a scalar field with canonical
action\ \eqref{canAction}.
Similarly, for tensor modes the Mukhanov variable is
\be
\mu \, \equiv \, ah \, ,
\ee
where $h$ is the amplitude of the polarization
tensor $h_{ij}$ (the two polarization states evolve independently
at linear order and obey the same equation of motion)
and the resulting equation of motion is
\begin{equation}
\label{EOMmuk}
 \mu_k''+\left(c_s^2k^2-\frac{a''}{a}\right)\mu_k = 0~.
\end{equation}
Alternatively, without the use of the Mukhanov variables, the equation of
motion for curvature and tensor perturbations
can be written as
\begin{align}
\label{EOMR}
 \mathcal{R}_k''+2\frac{z'}{z}\mathcal{R}_k'+c_s^2k^2\mathcal{R}_k& = 0~, \\
 \label{EOMh}
 h_k''+2\frac{a'}{a}h_k'+c_s^2k^2h_k& = 0~,
\end{align}
respectively.

Finally, let us introduce the scalar perturbation variable
\begin{equation}
 \zeta\equiv\Phi+\frac{2}{3}\frac{\Phi'+\mathcal{H}\Phi}{\mathcal{H}(1+w)}~,
\end{equation}
where $w\equiv P/\rho$ is the equation of state parameter ($P$ is the
pressure and $\rho$ is the energy density). On super-Hubble scales,
i.e.\ for $k\ll\mathcal{H}$, this variable is equivalent to the curvature
perturbation variable $\mathcal{R}_k$\ \cite{Bardeen:1980kt}. In other words,
$\mathcal{R}_k=\zeta_k$, and thus, throughout the rest of this paper,
we will use $\mathcal{R}_k$ and $\zeta_k$ interchangeably to denote
curvature perturbations on super-Hubble scales.

\section{Outline of the no-go conjecture}\label{sec:outlinenogo}

As explained in the introduction, a careful study of the evolution of
curvature perturbations and the production of non-Gaussianities during a
nonsingular bounce may lead to a ``no-go'' theorem, the impossibility of
obtaining a sufficiently small tensor-to-scalar ratio while maintaining a
bispectrum with an amplitude smaller than the current observational bounds.
In this section we will provide a qualitative analysis of this problem by
giving simple estimates of the tensor-to-scalar ratio and of the amplitude
of the bispectrum assuming that the curvature fluctuations undergo some
growth through the bounce phase. We first start by setting up the matter
bounce formalism.

\subsection{Fluctuations in the matter bounce}\label{sec:flucmatterbounce}

In the matter bounce, primordial quantum fluctuations originate on
sub-Hubble scales during a matter-dominated contracting phase and exit the
Hubble radius during this phase. The perturbations then remain on
super-Hubble scales as the universe contracts and passes through the bounce
phase, except for a very small time interval right at the bounce point (at
which time the Hubble radius goes to infinity). The fluctuations with
wavelength of cosmological interest today will then reenter the Hubble
radius in the standard radiation or matter-dominated expanding phases. If the
bounce is completely symmetric, then fluctuations which exit the Hubble
radius in the matter phase of contraction reenter the Hubble radius in the
matter phase of expansion. However, we expect the bounce to be asymmetric
and entropy to be generated during the bounce. In this case, the radiation
phase of expansion is longer than the radiation phase of contraction.

To understand the evolution of quantum fluctuations in a contracting
universe, one needs to determine the form of the variable $z$ and then
solve Eq.\ \eqref{EOMvk}. Using the Friedmann equations, the time
derivative of the Hubble parameter is given by
\begin{equation}\label{eq:Hdot}
 \dot H=-\frac{\dot\phi_0^2}{2}~,
\end{equation}
where a dot denotes a derivative with respect to cosmic time, $t$, and the
subscript $0$ indicates that we are referring to the background field.
Defining the parameter $\epsilon$,
\begin{equation}
 \epsilon\equiv-\frac{\dot H}{H^2}~,
\end{equation}
and using Eq.\ \eqref{eq:Hdot}, one finds
\begin{equation}
 z=a\frac{\dot\phi_0}{H}M_p\, = \, a \sqrt{2\epsilon} M_p~.
\end{equation}
It is straightforward to show from the Friedmann equations that
\begin{equation}
 \epsilon=\frac{3}{2}(1+w)~,
\end{equation}
so for a matter-dominated contracting universe with $w=0$, we have
$\epsilon=3/2$.
As a consequence, $z=a\sqrt{3}M_p$ and
\be
 \frac{z''}{z} \, = \, \frac{a''}{a}~,
\ee
and we conclude that the scalar and tensor fluctuations evolve in
exactly the same way. This is not true in general since $w$ can
vary in time. For example, in the case of inflationary cosmology,
we recognize $\epsilon$ as the slow-roll parameter and it is
time-dependent.

In a matter-dominated contracting universe, the scale factor scales as
$a\sim (-t)^{2/3}\sim\eta^2$, and since $c_s^2=1$ for a canonical scalar
field, the equation for the Sasaki-Mukhanov variable is
\begin{equation}
 v_k''+\left(k^2-\frac{2}{\eta^2}\right)v_k = 0~.
\end{equation}
On super-Hubble scales, the $k^2$ term is negligible, and so the solution reads
\begin{equation}
 v_k(\eta) \, = \, c_1\eta^2+c_2\eta^{-1}~.
\end{equation}
Using the fact that $v_k=z\zeta_k$, the first term yields $\zeta_k\sim\mathrm{constant}$,
but in a contracting universe, the second term is the dominant solution,
\begin{equation}
 \zeta_k\sim\eta^{-3}~,
\end{equation}
which implies that curvature perturbations grow in a contracting universe.
In fact, the growth rate is precisely the correct one to convert an initial
vacuum spectrum into a scale-invariant one (see e.g.\ \cite{Brandenberger:2010dk} for
a review).

\subsection{Bound from the tensor-to-scalar ratio}\label{sec:rmb}

The tensor-to-scalar ratio is defined as
\begin{equation}
 r\equiv\frac{\mathcal{P}_t(k_\ast)}{\mathcal{P}_{\zeta}(k_\ast)}~,
\end{equation}
where $k_\ast$ is the pivot scale which is used to parametrize the
power spectra for tensor and curvature perturbations. The
individual power spectra are defined by\ \cite{Liddle:1993fq}
\begin{align}
 \mathcal{P}_t(k)=&~2 \mathcal{P}_h(k)\equiv 2 \times 16 \pi \frac{k^3}{2\pi^2}|h_k|^2
 = 16 \pi \frac{k^3}{\pi^2}\frac{|\mu_k|^2}{a^2}~, \\
 \mathcal{P}_{\zeta}(k)\equiv&~\frac{k^3}{2\pi^2}|\zeta_k|^2=\frac{k^3}{2\pi^2}\frac{|v_k|^2}{z^2}~,
\end{align}
respectively. The factor of $2$ in the first step of the first line comes from the
two polarization states of gravitons and the factor of $16 \pi$ is a convention
reflecting the fact that it is $16 \pi M_p h$ which yields the canonical action
of a free scalar field in an expanding background\ \cite{Mukhanov:1990me}.

As we found in the previous subsection, $z = a\sqrt{3}M_p$ for the matter bounce, so
the scalar power spectrum becomes
\begin{equation}
 \mathcal{P}_{\zeta}(k)=\frac{k^3}{6\pi^2}\frac{|v_k|^2}{a^2 M_p^2} \, ,
\end{equation}
and furthermore, the tensor-to-scalar ratio becomes
\begin{equation}
 r \, = \, 96 \pi \left|\frac{\mu_{k_\ast}}{v_{k_\ast}}\right|^2 M_p^2\, .
\end{equation}
where the factor $M_p^2$ reflects the fact that we have defined $v_k$ to
have dimensions of mass, whereas $\mu_k$ is dimensionless.

Since $z''/z=a''/a$ for the matter bounce,
the evolution of scalar and tensor modes given by Eqs.\ \eqref{EOMvk}
and\ \eqref{EOMmuk}, respectively, will be identical. In addition, if
they originate from the same quantum vacuum,
then $v_k(\eta) = M_p \mu_k(\eta)$. Consequently, we find that $r = 96 \pi$.
If perturbations passed through the bounce unchanged, it would result
in $r = 96 \pi$  at the beginning of the standard big bang cosmology
phase which is three orders of magnitude larger than the current
observational upper bound.

To gain some intuition on the effect of passing through the bounce phase,
let us assume that curvature perturbations are enhanced by an amount
$\Delta\zeta_k$ through the bounce, i.e.
\begin{equation}
 \zeta_k(\eta_{B+})=\zeta_k(\eta_{B-})+\Delta\zeta_k~,
\end{equation}
where $\eta_{B\pm}$ denote the conformal time before ($-$) and after ($+$) the bounce. Then, the tensor-to-scalar
ratio measured after the bounce becomes
\begin{equation}
 r(\eta_{B+}) =
 96 \pi \left|\frac{h_{k_\ast}(\eta_{B+})}{\zeta_{k_\ast}(\eta_{B-})+\Delta\zeta_{k_\ast}}\right|^2~.
\end{equation}
Assuming that tensor modes remain constant through the bounce,
i.e.\ $h_k(\eta_{B-})=h_k(\eta_{B+})$, one finds that
\begin{equation}
 \left|1+\frac{\Delta\zeta_{k_\ast}}{\zeta_{k_\ast}(\eta_{B-})}\right|^2=\frac{r(\eta_{B-})}{r(\eta_{B+})}~.
 \label{zetaasrratio}
\end{equation}
Taking the value of the tensor-to-scalar ratio before the bounce to be
what we found earlier, i.e.\ $r(\eta_{B-}) = 96 \pi$, and demanding
that the tensor-to-scalar ratio is sufficiently suppressed after the
bounce so that it satisfies the observational bound
$r(\eta_{B+}) < 0.12$ (95\% CL from\ \cite{Ade:2015tva,Ade:2015lrj}),
we find that curvature perturbations must be sufficiently enhanced
during the bounce phase so that
\begin{equation}
 \left|1+\frac{\Delta\zeta_{k_\ast}}{\zeta_{k_\ast}(\eta_{B-})}\right| \gtrsim 50.1~,
\end{equation}
or using the triangle inequality,
\begin{equation}
\label{eq:constraintfromr}
 \left|\frac{\Delta\zeta_{k_\ast}}{\zeta_{k_\ast}(\eta_{B-})}\right| \gtrsim 49.1~.
\end{equation}

\subsection{Bound from the bispectrum}\label{sec:bispectrumintro}

The primordial bispectrum, $B_\zeta$, is defined in terms of the three-point function as
\begin{equation}
 \langle\zeta(\vec{k}_1)\zeta(\vec{k}_2)\zeta(\vec{k}_3)\rangle\equiv(2\pi)^3\delta^{(3)}(\vec{k}_1+\vec{k}_2+\vec{k}_3)B_\zeta(k_1,k_2,k_3)~,
\end{equation}
which we can rewrite as
\begin{equation}
 \langle\zeta(\vec{k}_1)\zeta(\vec{k}_2)\zeta(\vec{k}_3)\rangle=(2\pi)^7\delta^{(3)}\left(\sum_i\vec{k}_i\right)\frac{\mathcal{P}_\zeta^2}{\prod_ik_i^3}\mathcal{A}(k_1,k_2,k_3)~,
\end{equation}
where $k_i=|\vec{k}_i|$ and where the index $i$ runs from 1 to 3. The function $\mathcal{A}(k_1,k_2,k_3)$ is known as
the shape function and its amplitude defines the nonlinear parameter $f_{\mathrm{NL}}$ via
\begin{equation}
 f_{\mathrm{NL}}(k_1,k_2,k_3)=\frac{10}{3}\frac{\mathcal{A}(k_1,k_2,k_3)}{\sum_ik_i^3}~.
\end{equation}
Of particular interest is the local form of non-Gaussianities for which one of the three modes exits
the Hubble radius much earlier than the other two, i.e.\ $k_1\ll k_2=k_3$. For this case, one can write
\begin{equation}
 \zeta(\vec{x})=\zeta_g(\vec{x})+\frac{3}{5}f_{\mathrm{NL}}^{\mathrm{local}}\zeta_g(\vec{x})^2~,
\end{equation}
where $\zeta_g$ is the Gaussian part of $\zeta$.

In order to compute $f_{\mathrm{NL}}$, one must evaluate the
three-point function. To leading order in the interaction coupling
constant, the three-point function is related to the interaction
Lagrangian, $L_{\mathrm{int}}$, via\ \cite{Maldacena:2002vr}
\begin{equation} \label{basic}
 \langle\zeta(t,\vec{k}_1)\zeta(t,\vec{k}_2)\zeta(t,\vec{k}_3)\rangle=i\int_{t_i}^td\tilde t~
 \langle[\zeta(t,\vec{k}_1)\zeta(t,\vec{k}_2)\zeta(t,\vec{k}_3),L_{\mathrm{int}}(\tilde t)]\rangle~,
\end{equation}
where the square brackets denote the commutator and where $t_i$
denotes the initial time before which there is no non-Gaussianity.
The interaction Lagrangian is obtained by evaluating the action up to
third order in perturbation theory
\begin{equation}
 L_{\mathrm{int}}(t)=\int d^3\vec{x}~\mathcal{L}_3(t,\vec{x})~,
\end{equation}
and for a canonical scalar field, the Lagrangian density for $\zeta$ to
cubic order is given by\ \cite{Maldacena:2002vr}
\begin{align}
\label{L3canonical}
 \frac{\mathcal{L}_3}{M_p^2}&=\left(\epsilon^2-\frac{\epsilon^3}{2}\right)a^3\zeta\dot\zeta^2+\epsilon^2a\zeta(\partial\zeta)^2-2\epsilon^2a^3\dot\zeta(\partial\zeta)(\partial\chi)
 +\frac{\epsilon^3}{2}a^3\zeta(\partial_i\partial_j\chi)^2+f(\zeta)\frac{\delta\mathcal{L}_2}{\delta\zeta}~, \\
 f(\zeta)&=\frac{1}{4(aH)^2}(\partial\zeta)^2-\frac{1}{4(aH)^2}\partial^{-2}\partial_i\partial_j(\partial_i\zeta\partial_j\zeta)
 -\frac{1}{H}\zeta\dot\zeta-\frac{\epsilon}{2H}\partial_i\zeta\partial_i\chi+\frac{\epsilon}{2H}\partial^{-2}\partial_i\partial_j(\partial_i\chi\partial_j\zeta)~,
\end{align}
where $\partial^{-2}$ is the inverse Laplacian and where we define
$\chi\equiv\partial^{-2}\dot\zeta$.
Also, the equation of motion for $\zeta$ coming from the second
order perturbed Lagrangian density $\mathcal{L}_2$ is given by
\begin{equation}
 \frac{\delta\mathcal{L}_2}{\delta\zeta}=\frac{\partial}{\partial t}(az^2\dot\zeta)-\frac{c_s^2z^2}{a}\partial^2\zeta~.
\end{equation}

As we saw in Sec.\ \ref{sec:flucmatterbounce}, curvature perturbations
grow on super-Hubble scales during the matter-dominated contracting
phase until the bounce phase.
While on super-Hubble scales the spatial gradient terms are negligible, i.e.\
$\partial_i \zeta,\,\partial_i \chi \, \simeq \,0$,
the growth in $\zeta$ implies that the interaction Lagrangian is
dominated by
\begin{equation}
 \frac{\mathcal{L}_3}{M_p^2}\simeq\left(\epsilon^2-\frac{\epsilon^3}{2}\right)a^3\zeta\dot\zeta^2-\frac{1}{H}\zeta\dot\zeta\frac{\partial}{\partial t}(az^2\dot\zeta)~.
\end{equation}

As was first shown in\ \cite{Cai:2009fn}, the production of
non-Gaussianities on a comoving scale $k$ is dominated by the period
between when the scale crosses the Hubble radius in the phase of
matter contraction until the onset of the bounce phase, and the
resulting non-Gaussianities are of order
$f_{\mathrm{NL}}\sim\mathcal{O}(1)$. For example, for the local shape,
the authors of\ \cite{Cai:2009fn} found
$f_{\mathrm{NL}}^{\mathrm{local}} = -35/16$.

Following what was done in the previous subsection, let us now assume
that curvature perturbations grow during the bounce phase. For
simplicity, let us assume that they grow linearly in time with
constant rate
\begin{equation}
 \dot\zeta=\frac{\Delta\zeta}{\Delta t_B}~,
\end{equation}
where the duration of the bounce is given by
$\Delta t_B\equiv t_{B+}-t_{B-}$. Then, in the limit $k\rightarrow 0$
on super-Hubble scales, the contribution to the three-point function
coming from the bounce phase is schematically given by
\begin{equation}
 \langle\zeta(t_{B+})^3\rangle_{\mathrm{bounce}}\sim \frac{\zeta(t_{B+})^3}{M_p}\left(\frac{\Delta\zeta}{\Delta t_B}\right)^2\int_{t_{B-}}^{t_{B+}}dt
 ~a(t)^3\left[\epsilon(t)^2-\frac{\epsilon(t)^3}{2}\right]\left[\zeta(t_{B-})+\frac{\Delta\zeta}{\Delta t_B}(t-t_{B-})\right]~,
\end{equation}
and one expects that the dominant contribution to
$f_{\mathrm{NL}}$ that results from
evaluating the three-point function would scale as
\begin{equation}
 f_{\mathrm{NL}}\sim\frac{(\Delta\zeta)^2}{\Delta t_B}M_p^2~,
\end{equation}
plus terms of order $\Delta\zeta^1$
which would be subdominant for a large amplification $\Delta\zeta$.

We already see that a growth in the curvature perturbations during
the bounce, $\Delta\zeta$, would enhance $f_{\mathrm{NL}}$. From the
previous subsection, we expect $\Delta\zeta$ to have a lower bound to
match current observational bounds on $r$, and thus, we expect to find
a lower bound on the amount of non-Gaussianities that are produced
during the bounce phase. However, we cannot determine whether this
contribution will be significant to $f_{\mathrm{NL}}\sim\mathcal{O}(1)$
and whether the resulting lower bound will exceed current observational
bounds without going into the details of the calculation.

\subsection{The no-go theorem}\label{sec:thenogoconjecture}

Now, let us state our conjecture.
\begin{thm}
For quantum fluctuations originating from a matter-dominated
contracting universe, an upper bound on the tensor-to-scalar ratio ($r$)
is equivalent to a lower bound on the amplification of curvature perturbations ($\Delta\zeta/\zeta$)
which in turn is equivalent to a lower bound on the amount of primordial non-Gaussianities ($f_{\mathrm{NL}}$).
Furthermore, if the initial quantum vacuum is a canonical Bunch-Davies vacuum with $c_s=1$,
if the nonsingular bounce phase is due to a single NEC violating scalar field,
and if general relativity holds at all energy scales, then satisfying the current observational
upper bound on the tensor-to-scalar ratio cannot be done without contradicting the current observational upper bounds on
$f_{\mathrm{NL}}$ (and vice-versa).
\end{thm}
In the rest of this paper, we will give an example of realization of this conjecture.

\section{A brief review of single field bouncing cosmology}\label{sec:bounce}

In the context of Einstein gravity, matter which violates the null energy
condition must be introduced in order to obtain a cosmological bounce. A
simple toy model is quintom cosmology, i.e.\ a model in which a scalar
field with opposite sign in the action compared to a usual scalar field is
introduced, and it is arranged that this field comes to dominate late in
the contracting phase, thus yielding a nonsingular bounce\ \cite{Cai:2007qw}.
A specific realization of this can be obtained in the Lee-Wick theory
\cite{Cai:2008qw}. These models, however, suffer from a ghost instability
\cite{Cline:2003gs}. To avoid this instability (at least at the perturbative level)
one can make use of the ghost condensation mechanism\ \cite{Lin:2010pf} or the
Galileon construction\ \cite{Qiu:2011cy,Easson:2011zy}.\footnote{Alternative possibilities
of alleviating this instability may be achieved by considering various modified gravity
implementations such as models of extended $F(R)$ gravity\ \cite{Bamba:2013fha, Nojiri:2014zqa},
modified Gauss-Bonnet gravity\ \cite{Bamba:2014mya}, and torsion gravity
scenarios\ \cite{Cai:2011tc, Amoros:2013nxa}.}
These mechanisms involve a modified kinetic term in the action.

As mentioned in the introduction, bouncing models typically also suffer from
the anisotropy problem, and to mitigate this problem, one can build into the
scenario an Ekpyrotic phase of contraction which occurs at some point after
the matter phase of contraction. Specifically, one can use a single scalar
field with a kinetic term designed to yield a nonsingular bounce, and a
potential energy function with a negative potential over some range of field
values which is designed to yield Ekpyrotic contraction\ \cite{Cai:2012va}.
In this approach, a second scalar field with canonical kinetic term and with
quadratic potential can be used to represent the regular matter of the Universe
\cite{Cai:2013kja}. In this paper we will not consider the role which this
second scalar field may play (for some ideas see\ \cite{Cai:2011zx})
but only consider the field $\phi$ which generates the
Ekpyrotic contraction and the nonsingular bounce.

Throughout this paper, we assume only Einstein gravity plus matter.
Thus, the action is given by
\begin{equation}
\label{actionunperturbed}
 S = \int d^4x ~ \sqrt{-g} \left(-\frac{M_p^2}{2}R+\mathcal{L}_m\right)~,
\end{equation}
where $g$ is the determinant of the metric, $R$ is the Ricci scalar,
and $\mathcal{L}_m$ is the matter Lagrangian. We assume that the matter
content is dominated by only one scalar field ($\phi$) before reheating
(the energy density of matter created via reheating becomes only important
after the bounce phase \textendash\ see\ \cite{Quintin:2014oea}).
Thus, for the dynamics of
the matter-dominated contracting era and the bounce phase to be described
by second order equations of motion, we consider a Lagrangian of the most
general form\ \cite{Nicolis:2008in}
\begin{equation}
 \mathcal{L}_m \, = \, K(\phi,X) + G(\phi,X)\Box\phi + \mathcal{L}_4+\mathcal{L}_5~,
 \label{Lagrangian}
\end{equation}
where the kinetic variable $X$ is defined as
\begin{equation}
 X\equiv\frac{1}{2}g^{\mu\nu}\nabla_{\mu}\phi\nabla_{\nu}\phi~,
\end{equation}
and where the d'Alembertian operator is defined as
\begin{equation}
 \Box\phi\equiv g^{\mu\nu}\nabla_{\mu}\nabla_{\nu}\phi~.
\end{equation}
We do not write down the explicit form that $\mathcal{L}_4$ and
$\mathcal{L}_5$ can take here, but the key point is that they involve higher
order derivatives. If we assume that the energy scale at which the bounce
occurs is low enough so that higher order derivative terms in the Lagrangian
are negligible, then we can assume that
$\mathcal{L}_4,\,\mathcal{L}_5\approx 0$.

For the bounce to be nonsingular, the above Lagrangian must violate the
null energy condition (NEC) at high energies. To do so, we assume the first
term of the Lagrangian to have the form
\begin{equation}
 K(\phi,X) = M_p^2[1-g(\phi)]X+\beta X^2-V(\phi)~,
 \label{K-term}
\end{equation}
where $\beta$ is some positive constant.
We see from Eq.\ \eqref{K-term} that when $g(\phi)>1$, the sign of the
kinetic term is reversed and a ghost condensate which violates the NEC is
formed\ \cite{Lin:2010pf,Creminelli:2007aq,Buchbinder:2007ad,Qiu:2011cy,Easson:2011zy}.
For this reason, one typically chooses the function $g(\phi)$ to have the
form
\begin{equation}
 g(\phi) = \frac{2g_0}{e^{-\sqrt{2/p}\phi}+e^{b_g\sqrt{2/p}\phi}}~,
\end{equation}
where $p$ and $b_g$ are positive constants. As $\phi\rightarrow 0$ at the
bounce point, $g(\phi)\rightarrow g_0$, and the constant $g_0$ is naturally
chosen to be $g_0>1$ to allow the NEC violation. We can also see from the
form of $g(\phi)$ above that as $\phi$ goes away from 0 and as the kinetic
variable $X$ becomes small outside the bounce phase, $g(\phi)$ rapidly goes to 0 and the
Lagrangian recovers its canonical form.

The potential $V(\phi)$ can be chosen in order to obtain an Ekpyrotic phase
of contraction. This can be done by means of a potential which is negative
for small values of $|\phi|$, but which approaches $V = 0$ exponentially at
large positive and negative field values. Specifically, we have chosen the
potential
\be
V(\phi) \, = \, - \frac{2V_0}{e^{- \sqrt{2/q} \phi} + e^{b_V \sqrt{2/q} \phi}} \, ,
\ee
where $V_0$, $q$, and $b_V$ are positive constants. Without the second term in the
denominator, one obtains the potential postulated in the Ekpyrotic scenario
\cite{Khoury:2001wf}.

One can then parametrize the background evolution during the bounce phase as
follows. The Hubble parameter grows linearly in time, passing through zero at
the time $t = t_B$ (the bounce point),
\begin{equation}
 H(t) \, = \, \Upsilon(t-t_B)~,
\end{equation}
where $\Upsilon$ is a positive constant. The scale factor immediately follows,
\begin{equation}
 a(t) \, = \, a_Be^{\Upsilon(t-t_B)^2/2}~.
\end{equation}
Also, the scalar field evolves as
\begin{equation}\label{eq:dotphit}
 \dot\phi(t)=\dot\phi_Be^{-(t-t_B)^2/T^2}~.
\end{equation}
Since $a_B$ and $t_B$ can be arbitrarily redefined, we see that the parameters
which describe the bounce phase are $\Upsilon$, $\dot\phi_B$, $t_{B-}$
(or $t_{B+}$ assuming a symmetric bounce),
and $T$. First, $\Upsilon$ gives the growth rate of the Hubble
parameter. Second, $\dot\phi_B$ gives the maximal growth rate of the
scalar field. Third, $\Delta t_B/T$ gives the dimensionless duration of the
bounce. They can be related to the Lagrangian parameters via
(see\ \cite{Cai:2012va,Cai:2013kja})
\begin{align}\label{eq:dotphiB}
 \dot\phi_B&\simeq\sqrt{\frac{2(g_0-1)}{3\beta}}M_p~, \\
 T&\simeq\frac{H_{B+}}{\Upsilon}\sqrt{\frac{2}{\ln(\dot\phi_B^2/6H_{B+}^2)}}~,
\end{align}
where $H_{B+}=\Upsilon(t_{B+}-t_B)$.

Given the model we have discussed and the bounce solution which we have
given in parametric form, we will now follow the evolution of the curvature
fluctuation variable $\zeta$ through the nonsingular bounce phase.

\section{Evolution of curvature perturbations during the bounce}\label{sec:pert}

\begin{figure}
\includegraphics[scale=0.8]{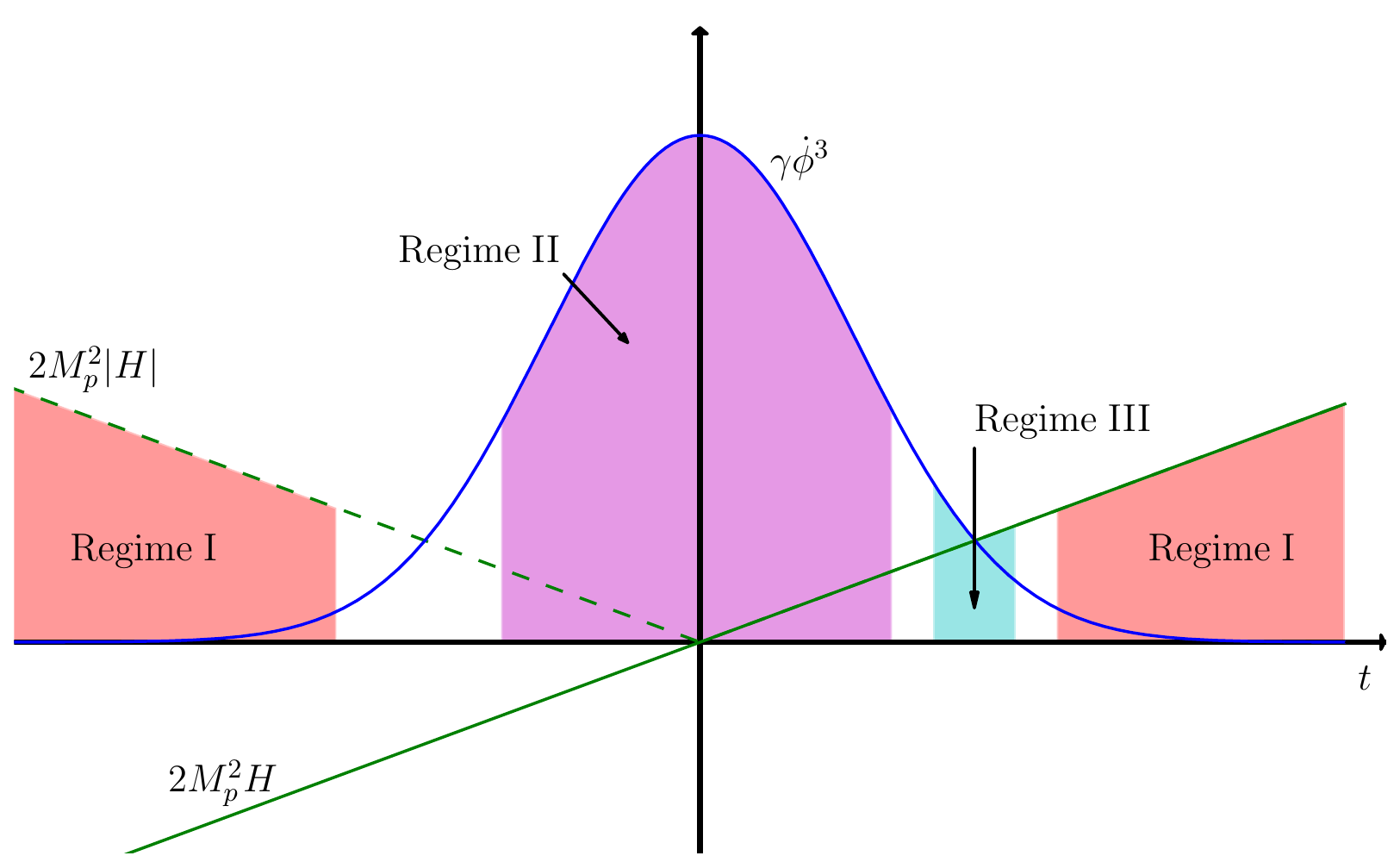}
\caption{Sketch of the different regimes in the bounce phase (not to scale).
The horizontal axis represents physical time. The green solid curve shows
$2M_pH(t)$ and the dashed version depicts its absolute value.
The bell-shaped blue curve represent $|G_{,X}(t)|\dot\phi^3(t)$, where we take
$G_{,X}=\gamma$ to be a positive constant for simplicity.
Regimes I, II, and III, defined by Eqs.\ \eqref{regimeI},\ \eqref{regimeII},
and\ \eqref{regimeIII}, are depicted by the pink, purple, and cyan regions,
respectively.}
\label{fig:sketch_bounce}
\end{figure}

As we saw in Sec.\ \ref{sec:cosmopert},
the equation of motion for curvature perturbations [Eq.\ \eqref{EOMR}] can
be written as
\begin{equation}
 \mathcal{R}_k''+\frac{(z^2)'}{z^2}\mathcal{R}_k'+c_s^2k^2\mathcal{R}_k=0~.
 \label{EOMR2}
\end{equation}
For a noncanonical Lagrangian of the form of Eq.\ \eqref{Lagrangian},
the variable $z$ and the sound speed are given by\ \cite{Cai:2012va}
\begin{align}
\label{z2}
 z^2=&~\frac{2M_p^4a^2\dot\phi^2\mathcal{P}}{(2M_p^2H-G_{,X}\dot\phi^3)^2}~, \\
 \label{cs2}
 c_s^2=&~\frac{1}{\mathcal{P}}\Big[K_{,X}+4H\dot\phi G_{,X}-\frac{G_{,X}^2\dot\phi^4}{2M_p^2}-2G_{,\phi}+G_{,X\phi}\dot\phi^2+(2G_{,X}+G_{,XX}\dot\phi^2)\ddot\phi\Big]~,
\end{align}
where a comma denotes a partial derivative and where we defined
\begin{equation}
 \mathcal{P}\equiv K_{,X}+\dot\phi^2 K_{,XX}+\frac{3}{2M_p^2}\dot\phi^4 G_{,X}^2+6H\dot\phi G_{,X}+3H\dot\phi^3 G_{,XX}-2G_{,\phi}-\dot\phi^2 G_{,\phi X}~.
\end{equation}
As explained in Sec.\ \ref{sec:flucmatterbounce}, the perturbation modes
that are of cosmological interest today were on super-Hubble scales during
the bounce phase (except in the immediate vicinity of the bounce point),
and thus we are most interested in the infrared (IR) regime of
Eq.\ \eqref{EOMR2}. In the limit $k\ll\mathcal{H}$, and recalling that
$\mathcal{R}_k$ and $\zeta_k$ are equivalent quantities in this limit,
the equation that we want to solve is
\begin{equation}
 \frac{d\zeta'}{d\eta}+\frac{(z^2)'}{z^2}\zeta'=0~,
 \label{EOMz}
\end{equation}
where we drop the $k$ index when it is clear that we are on super-Hubble
scales. It is obvious from the above equation that one solution is the
constant mode solution, $\zeta'=0$, that one expects on super-Hubble scales,
e.g.\ in inflation\ \cite{Bardeen:1983qw,Brandenberger:1983tg}
(see, however,\ \cite{Leach:2001zf}).
More generally, the solution to
Eq.\ \eqref{EOMz} can be written as
\begin{equation}
 \zeta'(\eta)=\zeta'(\eta_i)\frac{z^2(\eta_i)}{z^2(\eta)}~,
 \label{solzetaprime}
\end{equation}
where $\eta_i$ denotes the initial time where the initial conditions
are set. The evolution of $\zeta$ is thus governed by the evolution
of $z^2$, and we notice from the denominator of Eq.\ \eqref{z2} that
the evolution of $z^2$ has different regimes of interest:
\begin{align}
\label{regimeI}
 &\mathrm{Regime\ I}:\ 2M_p^2|H(t)|\gg |G_{,X}(t)|\dot\phi^3(t)~, \\
\label{regimeII}
 &\mathrm{Regime\ II}:\ 2M_p^2|H(t)|\ll |G_{,X}(t)|\dot\phi^3(t)~, \\
\label{regimeIII}
 &\mathrm{Regime\ III}:\ 2M_p^2H(t)\approx G_{,X}(t)\dot\phi^3(t)~.
\end{align}
We represent these different regimes in Fig.\ \ref{fig:sketch_bounce}, and
we explore the consequences of each regime in the following subsections.

\subsection{Evolution in Regime I}\label{sec:curv_regimeI}

When Eq.\ \eqref{regimeI} is valid, the expression for $z^2$ reduces to
\begin{equation}
 z^2 \, \simeq \, \frac{M_p^2a^2\dot\phi^2}{H^2}\left(\frac{1-g(\phi)}{2}\right)~.
\end{equation}
Since the bounce phase is defined by $g(\phi)>1$ and since $z^2$ must be positive to
avoid ghost instabilities, the model parameters must be chosen such that this
regime does not occur during the bounce phase. Outside the bouncing phase,
the equation of motion in Regime 1 reduces to the standard one.

\subsection{Evolution in Regime II}\label{sec:curv_bounce}

As the bounce point approaches, $H(t)$ goes to zero and we can expect
Eq.\ \eqref{regimeII} to be valid. To explore this regime, let us simplify
the treatment by setting
\begin{equation}
 G(\phi,X) \, = \, \gamma X
\end{equation}
for some positive constant $\gamma$, so the regime becomes
\begin{equation}
 2M_p^2|H(t)| \, \ll \, \gamma\dot\phi^3(t)~.
 \label{close_to_bounce}
\end{equation}
Using the parametrizations introduced in the previous section, this
condition can be rewritten as
\begin{equation}
 |\Delta t|e^{3(\Delta t)^2/T^2} \, \ll \, \frac{\gamma\dot\phi_B^3}{2M_p^2\Upsilon}~,
\end{equation}
where we defined $\Delta t\equiv t-t_B$. Since $\Delta t_B/T$ determines
the dimensionless duration of the bounce, remaining close to the bounce is
equivalent to demanding that $|\Delta t|/T\ll 1$.
In particular, if we demand that
\begin{equation}\label{eq:regimeIIcondition}
 |\Delta t| \, \ll \, \mathrm{min}\left\{\frac{T}{\sqrt{3}},\frac{\gamma\dot\phi_B^3}{2M_p^2\Upsilon}\right\}~,
\end{equation}
then it is ensured that we are in the regime set by
Eq.\ \eqref{close_to_bounce}.
Thus, the expression for $z^2$ given in
Eq.\ \eqref{z2} reduces to
\begin{equation}
 z^2(t) \, \simeq \, \frac{3\beta M_p^4}{\gamma^2}\frac{a^2(t)}{\dot\phi^2(t)}
 \label{z2_close_bounce}
\end{equation}
in this regime. In fact, there exists a time interval,
which we define as $[t_{\mathrm{amp}-},t_{\mathrm{amp}+}]$
with $t_{\mathrm{amp}\pm}\equiv t_B\pm\Delta t_{\mathrm{amp}}$,
where the above approximation for $z^2(t)$ is certainly valid.
We note that this expression is everywhere finite in that
interval, so the solution to
Eq.\ \eqref{EOMz} can be directly written as
\begin{equation}\label{eq:dotzetasol}
 \dot\zeta(t) \, = \, \dot\zeta(t_i)\frac{a(t_i)z^2(t_i)}{a(t)z^2(t)}~,
\end{equation}
where the initial condition must be taken in the interval,
i.e.\ $t_i\in[t_{\mathrm{amp}-},t_{\mathrm{amp}+}]$,
so logically we take $t_i=t_{\mathrm{amp}-}$.
Also, the solution will only be valid up to $t_{\mathrm{amp}+}$.
Inserting Eq.\ \eqref{z2_close_bounce} and using the parametrizations
introduced in the previous section, one finds
\begin{align}
 \zeta(t)\simeq&~\zeta(t_{\mathrm{amp}-})+\dot\zeta(t_{\mathrm{amp}-})\int_{t_{\mathrm{amp}-}}^td\tilde t~\left(\frac{a(t_{\mathrm{amp}-})}{a(\tilde t)}\right)^3\left(\frac{\dot\phi(\tilde t)}{\dot\phi(t_{\mathrm{amp}-})}\right)^2 \nonumber \\
 =&~\zeta(t_{\mathrm{amp}-})+\dot\zeta(t_{\mathrm{amp}-})\left(\frac{a(t_{\mathrm{amp}-})}{a_B}\right)^3\left(\frac{\dot\phi_B}{\dot\phi(t_{\mathrm{amp}-})}\right)^2\int_{t_{\mathrm{amp}-}}^{t}d\tilde t~\exp\left[-\left(\frac{2}{T^2}+\frac{3}{2}\Upsilon\right)(\tilde t-t_B)^2\right] \nonumber \\
 =&~\zeta(t_{\mathrm{amp}-})+\dot\zeta(t_{\mathrm{amp}-})\left(\frac{a(t_{\mathrm{amp}-})}{a_B}\right)^3\left(\frac{\dot\phi_B}{\dot\phi(t_{\mathrm{amp}-})}\right)^2T\sqrt{\frac{\pi}{8+6T^2\Upsilon}} \nonumber \\
  &\times\left[\mathrm{erf}\left(\frac{t-t_B}{T}\sqrt{2+\frac{3T^2\Upsilon}{2}}\right)-\mathrm{erf}\left(\frac{t_{\mathrm{amp}-}-t_B}{T}\sqrt{2+\frac{3T^2\Upsilon}{2}}\right)\right]~.
\end{align}
Close to the bounce point, the scale factor remains nearly constant,
so $a(t)\simeq a_B$. This implies that $\Upsilon(\Delta t)^2\ll 2$,
or in other words, that $H(t)\Delta t\ll\mathcal{O}(1)$.
We will assume this to be valid throughout the rest of this paper
whenever we are in the time interval $|\Delta t|\leq\Delta t_{\mathrm{amp}}$.
Therefore, the solution for $\zeta(t)$ reduces to
\begin{equation}\label{eq:solzetaRegimeII}
 \zeta(t)\simeq\zeta(t_{\mathrm{amp}-})+\dot\zeta(t_{\mathrm{amp}-})\left(\frac{\dot\phi_B}{\dot\phi(t_{\mathrm{amp}-})}\right)^2\frac{T\sqrt{2\pi}}{4}
  \left[\mathrm{erf}\left(\frac{t-t_B}{T}\sqrt{2}\right)-\mathrm{erf}\left(\frac{t_{\mathrm{amp}-}-t_B}{T}\sqrt{2}\right)\right]~.
\end{equation}
From the above solution, we see the constant mode and the growing mode.
Whether the constant or the growing mode is dominant
depends on many factors. For instance, the duration of this regime and
the growth rate will play a crucial role.
From the properties of the error function,
we note that the growing mode grows at most linearly in
time. Furthermore, the growth rate is maximal at the
bounce point $t_B$ and it is given by
$\dot\zeta_{\mathrm{max}}\simeq\dot\zeta(t_{\mathrm{amp}-})[\dot\phi_B/\dot\phi(t_{\mathrm{amp}-})]^2$.

\subsection{Evolution in Regime III}

One can notice from Eq.\ \eqref{solzetaprime} that if
$z^2\rightarrow\infty$, then $\zeta'\rightarrow 0$, and
curvature perturbations remain constant on super-Hubble scales.
One can see from Eq.\ \eqref{z2} that this happens at some physical time $t_s$
(or $\eta_s$ in conformal time) when
\begin{equation}
 2M_p^2H(t_s)=G_{,X}(t_s)\dot\phi^3(t_s)~.
\end{equation}
At this point, the equation of motion for the curvature perturbations becomes
singular, and furthermore, the Mukhanov variable
$v_k=z\mathcal{R}_k$ diverges. For this reason, the evolution of the
curvature perturbations has been explored in another gauge,
the harmonic gauge (first introduced in the context of cosmological perturbation theory in\ \cite{Xue:2013bva}),
where this singularity may disappear.
Using the harmonic gauge, it has been shown in\ \cite{Battarra:2014tga} that at $\eta_s$,
\begin{equation}
 \left.\frac{d\mathcal{R}_k}{d\eta}\right|_{\eta=\eta_s}=0~
\end{equation}
for all $k$ modes. Carefully dealing with the singular equation of motion
in the conformal Newtonian gauge,
one can find that in the IR limit, the solution in conformal time
close to the singular time $\eta_s$ is
(see Appendix\ \ref{sec:curv_pert_sing})
\begin{equation}
 \zeta(\eta)=\zeta(\eta_i)+\zeta'(\eta_i)\left(\frac{(\eta-\eta_s)^3+(\eta_s-\eta_i)^3}{3(\eta_s-\eta_i)^2}\right)~.
\end{equation}
This indicates that perturbations can grow before the singular point (coming from Regime II),
and that they could grow after the singular point (toward Regime I), but we saw that this
regime is not present in the bounce phase (Sec.\ \ref{sec:curv_regimeI}), so the bounce phase
will end shortly after the singular point $\eta_s$.

\subsection{Discussion}\label{sec:discussionzetabounce}

\begin{figure}
\includegraphics[scale=0.8]{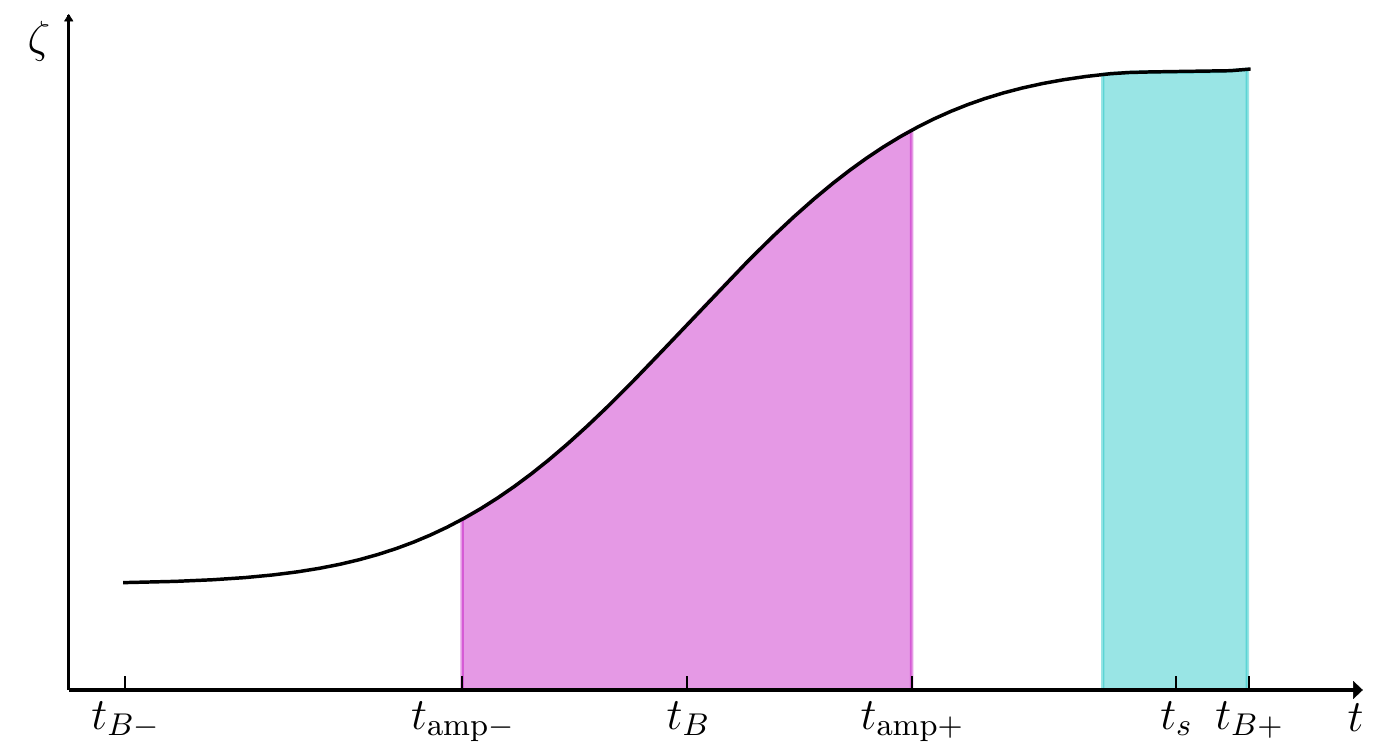}
\caption{Sketch of the evolution of curvature perturbation $\zeta$
on super-Hubble scales as a function of physical time $t$ (not to scale).
The beginning of the bounce phase, the bounce point, and the end of the
bounce phase are denoted by $t_{B-}$, $t_B$, and $t_{B+}$, respectively.
We defined $t_{\mathrm{amp}\pm}\equiv t_B\pm\Delta t_{\mathrm{amp}}$, and
$t_s$ is the time at which $z^2\rightarrow\infty$.
The purple region corresponds to Regime II of
Fig.\ \ref{fig:sketch_bounce}, where $\zeta$ grows at most linearly.
The cyan region corresponds to Regime III of
Fig.\ \ref{fig:sketch_bounce}, where $\zeta$ is almost constant.}
\label{fig:sketch_zeta}
\end{figure}

Let us summarize the evolution of curvature perturbations on super-Hubble
scales through the bounce phase. Figure\ \ref{fig:sketch_zeta} is a sketch of
the evolution of $\zeta$ according to the results found above.
If $\zeta$ enters the bounce phase with a nonvanishing time
derivative\footnote{If $\zeta$ enters the bounce phase with a vanishingly
small time derivative, then curvature perturbations will remain constant
throughout and exit the bounce phase unaffected.},
then we find that
curvature perturbations can grow at most linearly in some time interval
$[t_{\mathrm{amp}-},t_{\mathrm{amp}+}]$ and that the growth is maximal at
the bounce point. This happens in what we call Regime II.
We highlight this regime in purple in Fig.\ \ref{fig:sketch_zeta}.
Regime III follows Regime II at which point $z^2$ blows up and becomes
infinite at some time $t_s$.
At this point, curvature perturbations become constant, and the bounce phase
ends shortly after.
We highlight this regime in cyan in Fig.\ \ref{fig:sketch_zeta}.

In the end, the amplification that $\zeta$ receives is dominated by the
growth during the interval $[t_{\mathrm{amp}-},t_{\mathrm{amp}+}]$.
Between the beginning of the bounce phase and the beginning of the
amplification phase, we expect little growth of the curvature
perturbations, and so, the initial time derivative of $\zeta$
at the beginning of the amplification phase should be of the same order
as the time derivative of $\zeta$ at the beginning of the bounce phase,
which we expect to be small.
In fact, in the Ekpyrotic phase of contraction (where $w\gg 1$)
which precedes the bounce phase, the dominant mode of $\zeta$ is constant
in time while the second mode is decaying
(as shown in Appendix\ \ref{sec:pertoutsidebounce}).
Hence, the amplitude of $\zeta$ at the end of the period of Ekpyrotic contraction
is the same as the amplitude at the end of the matter phase of contraction (assuming
for a moment that there is no intermediate radiation phase).
Consequently, this could lead to a suppression of ${\dot \zeta}(t_{B-})$,
and hence to a suppression of the growth of $\zeta$ in the bounce phase
since, as we argued, $\dot\zeta(t_{B-})\simeq\dot\zeta(t_{\mathrm{amp}-})$.

The reason why we can match $\zeta$ and $\dot\zeta$ at the
end of the Ekpyrotic phase of contraction with the
beginning of the bounce phase comes from the matching conditions of
cosmological perturbations\ \cite{Cai:2008qw,Hwang:1991an,Deruelle:1995kd}.
These conditions impose that the gravitational potential $\Phi_k(\eta)$
and the modified curvature perturbation variable $\hat\zeta_k(\eta)$
are continuous across any transition (e.g.\ from the Ekpyrotic phase of
contraction to the bounce phase).
The variable $\hat\zeta_k$ is defined as\ \cite{Cai:2008qw}
\begin{equation}
 \hat\zeta_k\equiv\zeta_k+\frac{1}{3}c_s^2\left(\frac{k}{\mathcal{H}}\right)^2\Phi_k\left(1-\frac{\mathcal{H}'}{\mathcal{H}^2}\right)^{-1}~.
\end{equation}
On super-Hubble scales ($k\ll\mathcal{H}$), we note that the second term
of the above expression is suppressed, so $\hat\zeta_k\simeq\zeta_k$.
Thus, $\zeta_k$ must also be continuous across a transition.
That is why the values of $\zeta_k$ and $\dot\zeta_k$ at the end of the Ekpyrotic
phase of contraction are taken as the initial conditions of the bounce phase.

At this point, we note that the maximal growth rate
for $\zeta$ is given by
\begin{equation}\label{eq:zetadotmax}
 \dot\zeta_{\mathrm{max}}\simeq\dot\zeta(t_{B-})\left(\frac{\dot\phi_B}{\dot\phi(t_{\mathrm{amp}-})}\right)^2~,
\end{equation}
and that $\zeta$ grows at most linearly in time. Therefore,
one can say that
\begin{equation}
 \zeta(t_{\mathrm{amp}+})-\zeta(t_{\mathrm{amp}-})\lesssim\dot\zeta(t_{B-})\left(\frac{\dot\phi_B}{\dot\phi(t_{\mathrm{amp}-})}\right)^2(t_{\mathrm{amp}+}-t_{\mathrm{amp}-})~.
\end{equation}
Furthermore, since $\zeta$ receives essentially no amplification
outside the interval $[t_{\mathrm{amp}-},t_{\mathrm{amp}+}]$,
we can place an upper bound on the total growth
that curvature perturbations on super-Hubble scales receive from the
bounce phase,
\begin{equation}\label{eq:ubDeltazeta}
 \frac{\Delta\zeta}{\zeta(t_{B-})}\equiv\frac{\zeta(t_{B+})-\zeta(t_{B-})}{\zeta(t_{B-})}\lesssim\frac{\dot\zeta(t_{B-})}{\zeta(t_{B-})}\left(\frac{\dot\phi_B}{\dot\phi(t_{\mathrm{amp}-})}\right)^22\Delta t_{\mathrm{amp}}~,
\end{equation}
where we divide the growth $\Delta\zeta$ by the initial size
of $\zeta$ before the bounce to get a dimensionless quantity.

\subsection{Comparison with tensor modes}

We recall the equation of motion for tensor modes given by Eq.\ \eqref{EOMh},
which in the IR limit on super-Hubble scales reduces to
\begin{equation}
 h'' + 2\frac{a'}{a}h' \, = \, 0~.
\end{equation}
Once again, we drop the $k$ index when it is clear that the modes are in the IR limit.
Close to the bounce point, we recall that the scale factor is almost constant,
i.e.\ $a(\eta)\simeq a_B$. Thus, we are left with the equation $h''\simeq0$, and
consequently,
\be
h(\eta) \,\simeq \, h(\eta_i) + h'(\eta_i)(\eta-\eta_i) \, ,
\ee
or, equivalently,
\be \label{tensorgrowth}
h(t) \, \simeq \, h(t_i) + \dot h(t_i)(t-t_i) \, .
\ee

Thus, as in the case of curvature fluctuations in
Region II in the vicinity of the bounce point, there is a linearly
growing mode. Dimensional analysis, however, tells us that this growing
mode will not overwhelm the constant mode. The argument is as follows:
we can estimate ${\dot h(t_i)}$ to be of the order $M h(t_i)$, where $M$
is the mass scale at the bounce. On the other hand, we expect the
time interval of the bounce phase to be of the order $M^{-1}$, and hence
we expect the linearly growing term to be comparable at the end of the
bounce phase to the constant mode.

Comparing the coefficients of the linearly growing modes of the
curvature fluctuations and the tensor modes, i.e.\ Eq.\ \eqref{eq:zetadotmax}
and the coefficient of the growing mode in Eq.\ \eqref{tensorgrowth},
respectively, we see that it is the extra factor of
$[\dot\phi_B/\dot\phi(t_{\mathrm{amp}-})]^2$ in the coefficient of the scalar modes which
leads to the enhancement of the scalar power spectrum relative to the
tensor power spectrum.

\section{A comprehensive analysis of the production of primordial non-Gaussianities during the bounce phase}\label{sec:nong}

Now that we have identified the conditions under which the
tensor-to-scalar ratio can be suppressed,
we turn to the study of how the bispectrum
evolves during the bounce phase. We make use of the formalism
developed in\ \cite{Maldacena:2002vr} (see also
\cite{Chen:2010xka,Wang:2013zva}).

Our starting point is the expression\ \eqref{basic} for the three-point
function. From this expression it is clear that the bispectrum builds
up over time, which is to say that the three-point function after
the bounce equals the three-point function before the bounce plus
the result of integrating the right-hand side of\ \eqref{basic} over
the time interval of the bounce. From the form\ \eqref{L3canonical}
of the interaction Lagrangian it follows that the terms which dominate
the three-point function in the infrared are given by three powers of
$\zeta$ and two powers of its time derivative. As shown
explicitly in\ \cite{Cai:2009fn} in the computation of the three-point
function in the matter-dominated contracting phase, the absolute amplitude of
$\zeta$ cancels out in the definition of the shape function. Furthermore,
Cai {\it{et al.}}\ \cite{Cai:2009fn} show that the bispectrum at the end of the period of
matter contraction has an amplitude of the order $1$ with a shape which
is different from what is obtained in simple inflationary models.
Since the dominant mode of $\zeta$ is constant during the Ekpyrotic
phase of contraction, no additional contribution to the bispectrum is
generated during that phase. We have not computed the contribution
generated during a possible radiation phase of contraction between
the end of the matter period and beginning of the Ekpyrotic period.
This calculation could be done using the methods of\ \cite{Cai:2009fn}
and we would find again a contribution with amplitude of the order of
one and with a shape similar to that generated in the matter phase of
contraction and different from that in simple inflationary models, the
reason being that the same terms which dominate the bispectrum in the
matter phase will also dominate in the radiation phase, and they are
terms which are slow-roll suppressed during inflation.

Hence, we now turn to the evaluation of the contribution of the bouncing
phase to the three-point function. However, we must keep in mind that
the equations of\ \cite{Maldacena:2002vr}, in particular the third
order perturbed Lagrangian given by Eq.\ \eqref{L3canonical}, are only
valid for a canonical scalar field. We must generalize the analysis
to the case of the matter Lagrangian studied here (this generalization
will not affect the evolution of the three-point function outside of
the bounce phase because the extra terms which we derive below are
negligible except in the bounce phase).
This has already been done in the case of inflation for very general
Lagrangians (see, e.g.,\ \cite{Gao:2012ib,Choudhury:2015yna}).

For the Lagrangian given by Eq.\ \eqref{Lagrangian}, perturbations up to third order
in $\zeta$ yield the action
\begin{align}\label{eq:S3_3}
S_3=&\int d^4x~ \left( B_1 \left[\partial \zeta \partial \chi \partial ^2 \zeta- \zeta \partial _i \partial _j (\partial _i \zeta \partial_j \chi) \right]+ B_2 \dot{\zeta}^2 \partial ^2 \zeta \right. \nonumber \\
&+\left.B_3 \dot{\zeta} \partial \zeta \partial{\chi} +B_4 \zeta (\partial_i \partial_j\chi)^2 + B_5 \zeta (\partial \zeta)^2 +B_6 \dot{\zeta} ^3 +B_7\zeta\dot{\zeta} ^2-2f(\zeta) \frac{\delta {\cal{L}}_2}{\delta \zeta} \right)~,
\end{align}
where
\begin{equation}\label{eq:f}
f(\zeta)=\frac{A_{20} a^2}{4 M_p^2} \left[ (\partial \zeta)^2 -\partial ^{-2} \partial _i  \partial_ j (\partial _i \zeta \partial_j \zeta) \right] + \frac{A_{18} a^2}{M_p^2} \left[ \partial \zeta \partial \chi - \partial ^{-2} \partial _i \partial _j (\partial _i \zeta \partial_j \chi) \right] - \frac{2 A_4 a^3 - C_1}{2 z^2 c_s^2} a \zeta \dot{\zeta}~.
\end{equation}
The derivation of this action and the form of the functions $A_n$, $B_n$, and $C_n$ ($n=1,...$)
can be found in Appendix\ \ref{sec:thirdorderaction}.
As expected, this action is equivalent to the action given by Eq.\ \eqref{L3canonical} in the limit
where the Lagrangian\ \eqref{Lagrangian} is canonical in a matter-dominated contracting universe.
This is shown in Appendix\ \ref{sec:S3limitmatter}.

In order to cancel the last term in Eq.\ \eqref{eq:S3_3},
we make a field redefinition in Fourier space
$\zeta(\eta,\vec{k}) \rightarrow \zeta(\eta,\vec{k})-f(\eta,\vec{k})$
in the third order Lagrangian.
This way, there will be two contributions to the three-point function.
The first part of the three-point function is the third order Lagrangian
without the last term and the second part is related to the field
redefinition terms where $\zeta(\eta,\vec{k})$ is replaced by
$f(\eta,\vec{k})$. Using the Lagrangian formalism, we note that in Fourier space, we can
canonically express the modes $\zeta(\eta,\vec{k})$ as follows,
\begin{equation}
\label{modeexpansion}
 \zeta(\eta,\vec{k})=\zeta_k(\eta)a_{\vec{k}}^{\dag}+\zeta_k^*(\eta)a_{-\vec{k}}~,
\end{equation}
where $a_{\vec{k}}|0\rangle=0$, so $a_{\vec{k}}$ is the annihilation operator, and $a_{\vec{k}}^{\dag}$ is the respective creation operator.
Then, if we consider the interaction picture, the three-point function
to leading order in the interaction coupling constant is given by
\begin{equation}
\label{3pfuncint}
 \langle\zeta(\eta,\vec{k}_1)\zeta(\eta,\vec{k}_2)\zeta(\eta,\vec{k}_3)\rangle_{\mathrm{int}}=i\int_{\eta_i}^\eta d\tilde\eta~
 \langle[\zeta(\eta,\vec{k}_1)\zeta(\eta,\vec{k}_2)\zeta(\eta,\vec{k}_3),L_{\mathrm{int}}(\tilde\eta)]\rangle~,
\end{equation}
where $\eta_i$ corresponds to the initial time before which there is no
non-Gaussianity. Also, $L_{\mathrm{int}}$ is associated with the third
order action\ \eqref{eq:S3_3} without its last term.

Here, we are interested in the production of
non-Gaussianities during the bounce phase, so we consider the initial
time to be the beginning of the bounce phase and we consider the end
time at which the three-point function is evaluated to be the end of
the bounce phase. However, as we saw in the previous section,
curvature perturbations are nearly constant, and hence do not
contribute to the three-point function, except during the small time
interval $[\eta_{\mathrm{amp}-},\eta_{\mathrm{amp}+}]$ where $\zeta$
grows. Thus, the integration bounds are taken to be from
$\eta_{\mathrm{amp}-}$ to $\eta_{\mathrm{amp}+}$,
and the evolution of the curvature perturbations is taken to be
\begin{equation}
\label{eq:zetabounceNL}
 \zeta_k(\eta)=\zeta_k^\mathrm{m}(\eta_{B-})+\zeta_k^{\mathrm{m}\prime}(\eta_{B-})\left(\frac{\phi_B'}{\phi'(\eta_{\mathrm{amp}-})}\right)^2(\eta-\eta_{\mathrm{amp}-})~.
\end{equation}
The above expression follows from taking the maximal linear growth rate
given by Eq.\ \eqref{eq:zetadotmax} throughout the amplification
interval $[\eta_{\mathrm{amp}-},\eta_{\mathrm{amp}+}]$.
This expression slightly underestimates $\zeta_k$ for
$\eta_{\mathrm{amp}-}\leq\eta<\eta_B$ and
slightly overestimates $\zeta_k$ for
$\eta_B<\eta\leq\eta_{\mathrm{amp}+}$
but it is a good approximation on average over the small
interval $[\eta_{\mathrm{amp}-},\eta_{\mathrm{amp}+}]$.

We recall that curvature perturbations
are more or less constant during the Ekpyrotic
phase of contraction that precedes the bounce phase.
Therefore, it is natural to take the end conditions
of the matter-dominated phase of contraction
as the initial conditions of the bounce phase.
As shown in Sec.\ \ref{sec:discussionzetabounce},
$\zeta_k$ and $\zeta_k'$ must be continuous
across any transition on super-Hubble scales.
Hence for the initial conditions of the bounce phase,
we put the superscript ``m'' which
denotes the matter
bounce solution\ \cite{Cai:2009fn}
\begin{equation}\label{eq:zetam}
 \zeta_k^\mathrm{m}(\eta)=\frac{iAe^{ik(\eta-\tilde\eta_{B-})}[1-ik(\eta-\tilde\eta_{B-})]}{\sqrt{2k^3}(\eta-\tilde\eta_{B-})^3}~,
\end{equation}
where $\tilde\eta_{B-}$ is the conformal time at the singularity
if the matter-dominated contracting phase were to continue
to arbitrary densities (i.e.\ without NEC violating matter).
Also, $A$ is a normalization constant which is determined from the quantum vacuum
condition at Hubble radius crossing in the contracting phase,
and it is found to be
\begin{equation}\label{eq:Anorm}
 A=\frac{(\Delta\eta_{B-})^2}{\sqrt{3}a_BM_p}~,
\end{equation}
where $\Delta\eta_{B-}\equiv\tilde\eta_{B-}-\eta_{B-}$.

Let us comment on the wave number dependence of
Eq.\ \eqref{eq:zetabounceNL}. We first solved the equation of motion in
the bouncing phase in the limit where $k\ll\mathcal{H}$ to
$0^{\mathrm{th}}$ order. Then, matching the solution at the beginning
of the bouncing phase with the one at the end of the matter contraction phase, we
introduced some wave number dependence since the solution in the matter
contraction phase has higher order terms in $k/\mathcal{H}$. Thus, one may
worry that obtaining the correct $k$-dependent solution in the bounce
phase up to leading order requires one to solve the full $k$-dependent
equation of motion. However, we note that we will be interested in the
IR limit again when evaluating the three-point function. Thus, any $k$
dependence not included in the above solution is suppressed during the
bounce phase as long as $k$ remains much smaller than the largest
energy scale attained during the bounce, i.e.\ as long as
$k\ll\mathcal{H}_{B-},\mathcal{H}_{B+}$,
and as long as the corresponding wavelength of the
fluctuations remains much larger than the bounce length scale, which
can be reformulated as $k\ll(\Delta\eta_B)^{-1}$.

Substituting the interaction Lagrangian $L_{\mathrm{int}}$ associated with the action\ \eqref{eq:S3_3} without
its last term into Eq.\ \eqref{3pfuncint} and using
Eq.\ \eqref{modeexpansion}, we find
\begin{align}\label{eq:3ptfuncLcontrib}
\langle\zeta(\vec{k}_1)\zeta(\vec{k}_2)\zeta(\vec{k}_3)\rangle_{\mathrm{int}}=&~(2\pi)^3\delta^{(3)}\left(\sum_{i=1}^3\vec{k}_i\right)\zeta_{k_1}^*(\eta_{+})\zeta_{k_2}^*(\eta_{+})\zeta_{k_3}^*(\eta_{+}) \nonumber \\
 &\times i\int_{\eta_{-}}^{\eta_{+}}d\eta\left[\frac{B_1(\eta)z^2(\eta)}{M_p^2}\left(\frac{\vec{k}_1\cdot \vec{k}_3}{k_3^2} k_2^2 +\frac{[\vec{k}_2\cdot (\vec{k}_2-\vec{k}_3)][\vec{k}_3\cdot (\vec{k}_2+\vec{k}_3)]}{k_3 ^2}\right)\zeta_{k_1}(\eta)\zeta_{k_2}(\eta)\zeta_{k_3}'(\eta)\right. \nonumber \\
 &+\left(\frac{B_2(\eta)}{a_B} k_1^2 +\frac{B_3(\eta)z^2(\eta)}{M_p^2 a_B}\frac{\vec{k}_1\cdot\vec{k}_3}{k_3^2}+\frac{B_4(\eta)z^4(\eta)}{M_p^4 a_B}\frac{(\vec{k}_2\cdot\vec{k}_3)^2}{k_3^2 k_2^2}+\frac{B_7(\eta)}{a_B}\right)\zeta_{k_1}(\eta)\zeta_{k_2}'(\eta)\zeta_{k_3}'(\eta) \nonumber \\
 &+\left.B_5(\eta)a_B(\vec{k}_1\cdot\vec{k}_2)\zeta_{k_1}(\eta)\zeta_{k_2}(\eta)\zeta_{k_3}(\eta)+\left(\frac{B_6(\eta)}{a_B^2}\right)\zeta_{k_1}'(\eta)\zeta_{k_2}'(\eta)\zeta_{k_3}'(\eta)\right]+(5\ \mathrm{permutations})~.
\end{align}
Moreover, the contribution from the field redefinition is
\begin{align}\label{eq:3ptfuncfieldredef}
-\langle\zeta(\vec{k}_1)\zeta(\vec{k}_2)f(\vec{k}_3)\rangle_{\mathrm{redef}}=&~(2\pi)^3\delta^{(3)}\left(\sum_{i=1}^3\vec{k}_i\right) \nonumber \\
&\times\left[\frac{A_{20}(\eta_{+})a_B^2}{4 M_p^2}\left(-\vec{k}_1\cdot (\vec{k}_3-\vec{k}_1)+\frac{(\vec{k}_1\cdot\vec{k}_3)[(\vec{k}_3-\vec{k}_1)\cdot\vec{k}_3]}{k_3^2}\right) |\zeta_{k_1}(\eta_{+})|^2 |\zeta_{k_2}(\eta_{+})|^2\right. \nonumber \\
&-\frac{A_{18}(\eta_{+})a_Bz^2(\eta_{+})}{M_p^4}\left(\frac{\vec{k}_1\cdot (\vec{k}_3-\vec{k}_1)}{k_1^2}-\frac{(\vec{k}_1\cdot\vec{k}_3)[(\vec{k}_3-\vec{k}_1)\cdot\vec{k}_3]}{k_1^2 k_3^2}\right)\zeta_{k_1}'(\eta_{+})\zeta_{k_1}^*(\eta_{+})|\zeta_{k_2}(\eta_{+})|^2 \nonumber \\
&+\left.\left(\frac{2A_4(\eta_{+})a_B^3-C_1(\eta_{+})}{2z^2(\eta_{+})c_s^2}\right) \zeta_{k_1}'(\eta_{+})\zeta_{k_1}^*(\eta_{+})|\zeta_{k_2}(\eta_{+})|^2\right]+(5\ \mathrm{permutations})~.
\end{align}
The permutations that we refer to are over the $\vec{k}_i$ vectors for $i = 1, 2, 3$.
We note that, to simplify the notation, we set $\eta_{\pm}\equiv\eta_{\mathrm{amp}\pm}$.
The general form of the full three-point function can be expressed as
\begin{equation}\label{eq:3ptfunc}
 \langle\zeta(\vec{k}_1)\zeta(\vec{k}_2)\zeta(\vec{k}_3)\rangle=\langle\zeta(\vec{k}_1)\zeta(\vec{k}_2)\zeta(\vec{k}_3)\rangle_{\mathrm{int}}
 +\langle\zeta(\vec{k}_1)\zeta(\vec{k}_2)f(\vec{k}_3)\rangle_{\mathrm{redef}}
 =(2\pi)^7\delta^{(3)}\left(\sum_i\vec{k}_i\right)\frac{\mathcal{P}_\zeta^2}{\prod_ik_i^3}\mathcal{A}(k_1,k_2,k_3)~,
\end{equation}
and so, if we substitute Eqs.\ \eqref{eq:3ptfuncLcontrib}
and\ \eqref{eq:3ptfuncfieldredef} into the above,
we find the shape function to be given by
\begin{align}
 \mathcal{A}(k_1,k_2,k_3)=&~\frac{k_3^3\zeta^*_{k_1}(\eta_{+})\zeta^*_{k_2}(\eta_{+})\zeta^*_{k_3}(\eta_{+})}{4 \zeta^*_{k_1}(\eta_{+})\zeta^*_{k_2}(\eta_{+})\zeta_{k_1}(\eta_{+})\zeta_{k_2}(\eta_{+})} \nonumber \\
 &\times i\int_{\eta_{-}}^{\eta_{+}}d\eta\left[B_1(\eta)z^2(\eta)\left(\frac{\vec{k}_1\cdot\vec{k}_3}{k_3^2}k_2^2-\frac{[\vec{k}_2\cdot(\vec{k}_2+\vec{k}_3)][\vec{k}_3\cdot(\vec{k}_2+\vec{k}_3)]}{k_3^2}\right)\zeta_{k_1}(\eta)\zeta_{k_2}(\eta)\zeta_{k_3}'(\eta)\right. \nonumber \\
 &+\left(\frac{B_2(\eta)}{a_B}k_1^2 +\frac{B_3(\eta)z^2(\eta)}{M_p^2a_B}\frac{\vec{k}_1\cdot\vec{k}_3}{k_3^2}+\frac{B_4(\eta)z^4(\eta)}{M_p^4a_B}\frac{(\vec{k}_2\cdot\vec{k}_3)^2}{k_3^2k_2^2}+\frac{B_7(\eta)}{a_B}\right)\zeta_{k_1}(\eta)\zeta_{k_2}'(\eta)\zeta_{k_3}'(\eta) \nonumber \\
 &+\left.B_5(\eta)a_B(\vec{k}_1\cdot\vec{k}_2)\zeta_{k_1}(\eta)\zeta_{k_2}(\eta)\zeta_{k_3}(\eta)+\frac{B_6(\eta)}{a_B^2}\zeta_{k_1}'(\eta)\zeta_{k_2}'(\eta)\zeta_{k_3}'(\eta)\right] \nonumber \\
 &+\frac{A_{20}(\eta_+)a_B^2k_3^3}{4M_p^2}\left(-\vec{k}_1\cdot(\vec{k}_3-\vec{k}_1)+\frac{(\vec{k}_1\cdot\vec{k}_3)[(\vec{k}_3-\vec{k}_1)\cdot\vec{k}_3]}{k_3^2}\right) \nonumber \\
 &-\frac{A_{18}(\eta_+)a_Bz^2(\eta_+)k_3^3}{M_p^4}\left(\frac{\vec{k}_1\cdot(\vec{k}_3-\vec{k}_1)}{k_1^2}-\frac{(\vec{k}_1\cdot\vec{k}_3)[(\vec{k}_3-\vec{k}_1)\cdot\vec{k}_3]}{k_1^2k_3^2}\right)\frac{\zeta_{k_1}'(\eta_+)\zeta_{k_1}^*(\eta_+)}{|\zeta_{k_1}(\eta_+)|^2} \nonumber \\
 &+k_3^3\left(\frac{2A_4(\eta_+)a_B^3-C_1(\eta_+)}{2z^2(\eta_+)c_s^2}\right)\frac{\zeta_{k_1}'(\eta_+)\zeta_{k_1}^*(\eta_+)}{|\zeta_{k_1}(\eta_+)|^2}+(5\ \mathrm{permutations})~.
 \label{shapefunctionunevaluated}
\end{align}
At this point, we should note that the contributions coming from the
terms with coefficients $B_1$, $B_2$, $B_5$, and $A_{20}$ are of order
$\mathcal{O}(k^5)$, and consequently, these terms are vanishingly small
compared to other terms, which are of order $\mathcal{O}(k^3)$, on
super-Hubble scales. Therefore, the three main contributions to the
shape function are the $\zeta\zeta'^2$ term, the $\zeta'^3$ term, and
the field redefinition term.
We evaluate each of these terms
separately in Appendix\ \ref{sec:shape_function}
and we find the general expression for the shape function after
the bounce phase [see Eq.\ \eqref{shapefunctionfull}].

Three important forms of non-Gaussianity in cosmological observations
are the local form, the equilateral form, and the orthogonal form.
The local form of non-Gaussianity requires that one of the three
momentum modes exits the Hubble radius much earlier than the other two,
i.e. $k_1 \ll k_2=k_3$. Evaluating the shape
function\ \eqref{shapefunctionfull} in this limit yields
\begin{align}
f_{\mathrm{NL}}^{\mathrm{local}}=&~\frac{10}{3}\frac{{\cal{A}}(k_1 \ll k_2=k_3)}{\sum_{i}k_i^3} \nonumber \\
  \simeq&~\frac{10}{3}\left[\frac{-A^2}{8 a_B \Delta\eta_{B-}^4} \left(\frac{\phi'_B}{\phi'(\eta_-)}\right)^2
    \left(\frac{B_4(\eta_B)z_B^4}{M_p^4} + B_7(\eta_B)\right)
    -\frac{3 A^2}{8 \Delta\eta_{B-}^4 \Delta\eta_{\mathrm{amp}}} \left(\frac{\phi'_B}{\phi'(\eta_-)}\right)^2
    \frac{B_6(\eta_B)}{a_B^2} \right. \nonumber \\
    &\left. + \frac{A_{18}(\eta_{+})a_Bz^2(\eta_{+})}{8 M_p^4 \Delta\eta_{\mathrm{amp}}}
      +\frac{2 A_4(\eta_{+}) a_B^3 - C_1(\eta_{+})}{8 z^2(\eta_{+}) c_s^2 \Delta\eta_{\mathrm{amp}}}\right]~.
\end{align}
The equilateral form of non-Gaussianity requires that $k_1=k_2=k_3$,
so one finds
\begin{align}
f_{\mathrm{NL}}^{\mathrm{equil}}=&~\frac{10}{3}\frac{{\cal{A}}(k_1=k_2=k_3)}{\sum_{i}k_i^3} \nonumber \\
  \simeq&~\frac{10}{3}\left[\frac{A^2}{16 a_B \Delta\eta_{B-}^4} \left(\frac{\phi'_B}{\phi'(\eta_-)}\right)^2
  \left(\frac{B_3(\eta_B) z_B^2}{M_p^2} - \frac{B_4(\eta_B) z_B^4}{2 M_p^4} - 2 B_7(\eta_B)\right)
  -\frac{3 A^2}{8 \Delta\eta_{B-}^4 \Delta\eta_{\mathrm{amp}}} \left(\frac{\phi'_B}{\phi'(\eta_-)}\right)^2
    \frac{B_6(\eta_B)}{a_B^2} \right. \nonumber \\
    &\left. + \frac{3 A_{18}(\eta_{+}) a_B z^2(\eta_{+})}{16 M_p^4 \Delta\eta_{\mathrm{amp}}}
    +\frac{2 A_4(\eta_{+}) a_B^3 - C_1(\eta_{+})}{8 z^2(\eta_{+}) c_s^2 \Delta\eta_{\mathrm{amp}}}\right]~.
\end{align}
Finally, the orthogonal form of non-Gaussianity requires that $k_1=\sqrt{k_2^2+k_3^2}=\sqrt{2} k$,
so one finds
\begin{align}
f_{\mathrm{NL}}^{\mathrm{ortho}}=&~\frac{10}{3}\frac{{\cal{A}}(k_1=\sqrt{k_2^2+k_3^2}=\sqrt{2} k)}{\sum_{i} k_i^3} \nonumber \\
  \simeq&~\frac{10}{3}\left[\frac{A^2}{16 a_B \Delta\eta_{B-}^4} \left(\frac{\phi'_B}{\phi'(\eta_-)}\right)^2
  \left((4-3\sqrt{2}) \frac{B_3(\eta_B) z_B^2}{M_p^2}+(4-2\sqrt{2})\frac{B_4(\eta_B) z_B^4}{M_p^4}-2B_7(\eta_B)\right)\right. \nonumber \\
  &\left. -\frac{3 A^2}{8 \Delta\eta_{B-}^4 \Delta\eta_{\mathrm{amp}}} \left(\frac{\phi'_B}{\phi'(\eta_-)}\right)^2 \frac{B_6(\eta_B)}{a_B^2}
  +(1+\sqrt{2})\frac{A_{18}(\eta_{+}) a_B z^2(\eta_{+})}{16 M_p^4 \Delta\eta_{\mathrm{amp}}}
  +\frac{2 A_4(\eta_{+}) a_B^3 - C_1(\eta_{+})}{8 z^2(\eta_{+}) c_s^2 \Delta\eta_{\mathrm{amp}}}\right]~.
\end{align}

Substituting in some values for the model parameters
($\Upsilon$, $T$, $\phi'_B$, $\gamma$, etc., introduced
in Sec.\ \ref{sec:bounce}) would yield specific
numbers for the amount of non-Gaussianities that has
been produced during the bounce. However, instead of
giving exact values now, we will try to constrain
the parameter space from observations.
This is what we do in the next section.

\section{Combination of the observational bounds on non-Gaussianities and on the tensor-to-scalar ratio}\label{sec:constraints}

Let first rewrite the expression for
$f_{\mathrm{NL}}^{\mathrm{local}}$
using Eqs.\ \eqref{eq:Anorm},\ \eqref{eq:B4sf},
\eqref{eq:B7sf},\ \eqref{eq:B6sf},\ \eqref{eq:A18sf},
and\ \eqref{eq:A4C1sf},
\begin{align}
 f_{\mathrm{NL}}^{\mathrm{local}}\simeq&
 -\frac{5}{12\gamma^4\phi_B'^8}
 \left(3a_B^2\beta\gamma^2M_p^2\phi_B'^6+3\gamma^4\phi_B'^8+12a_B^8\beta M_p^6\Upsilon+16a_B^6\gamma^2M_p^4\phi_B'^2\Upsilon\right)
 \left(\frac{\phi_B'}{\phi'(\eta_-)}\right)^2 \nonumber \\
 &-\frac{25a_B^2M_p^2}{6\gamma^3\phi_B'^5}
 \left(2a_B^2\beta M_p^2+3\gamma^2\phi_B'^2\right)
 \frac{1}{\Delta\eta_{\mathrm{amp}}} \left(\frac{\phi_B'}{\phi'(\eta_-)}\right)^2
 +\frac{5a_B^4\beta M_p^4}{4\gamma^3\phi_B'^5}
 \frac{1}{\Delta\eta_{\mathrm{amp}}} \left(\frac{\phi_B'}{\phi'(\eta_-)}\right)^5 \nonumber \\
 &+\frac{5a_B^2M_p^2}{\gamma\phi_B'^3}
 \frac{1}{\Delta\eta_{\mathrm{amp}}} \left(\frac{\phi_B'}{\phi'(\eta_-)}\right)^3
 +\frac{10\gamma}{3\beta\phi_B'}
 \frac{1}{\Delta\eta_{\mathrm{amp}}} \left(\frac{\phi_B'}{\phi'(\eta_-)}\right)~.
\end{align}
The equilateral and orthogonal
$f_{\mathrm{NL}}$ have similar expressions, only with
different coefficients.

At this point, we do not want to insert
specific values for the model parameters.
Yet, in order to have a healthy bounce,
i.e.\ one that yields a bounce
free of ghost instabilities,
we expect the model parameters to
lie in specific regimes.
From\ \cite{Cai:2012va,Cai:2013kja,Cai:2014xxa},
we expect that $\phi_B'<a_BM_p$,
$\Upsilon\ll M_p^2$,
$\beta\sim\mathcal{O}(1)$,
and $\gamma\ll 1$.
Also, from Eq.\ \eqref{eq:dotphit},
it is obvious that $\phi_B'/\phi'(\eta_-)>1$.
Therefore, keeping only dominant terms, the
expression for $f_{\mathrm{NL}}^{\mathrm{local}}$
reduces to
\begin{equation}
 f_{\mathrm{NL}}^{\mathrm{local}}\simeq
 \frac{5a_B^4\beta M_p^4}{\gamma^3\phi_B'^5}
 \left(\frac{\phi_B'}{\phi'(\eta_-)}\right)^2
 \left[-\frac{a_B^4M_p^2\Upsilon}{\gamma\phi_B'^3}-\frac{5}{3\Delta\eta_{\mathrm{amp}}}+\frac{1}{4\Delta\eta_{\mathrm{amp}}}
 \left(\frac{\phi_B'}{\phi'(\eta_-)}\right)^3\right]~.
\end{equation}
In the square bracket, the three terms come from
$\mathcal{A}_{\zeta\zeta'^2}$,
$\mathcal{A}_{\zeta'^3}$,
and $\mathcal{A}_{\mathrm{redef}}$,
respectively. However, let us recall from
Appendix\ \ref{sec:shape_function} that
the results for $\mathcal{A}_{\zeta\zeta'^2}$
and $\mathcal{A}_{\zeta'^3}$ were actually
upper bounds in absolute value.
Since we expect that
$\Delta\eta_{\mathrm{amp}}\sim\mathcal{O}(1/a_BM_p)$
from Eq.\ \eqref{eq:regimeIIcondition},
it results that the field redefinition term
is dominant over the other ones,
just like in the regular matter bounce\ \cite{Cai:2009fn},
so we can write
\begin{equation}\label{eq:fNLlocalsimpl1}
 f_{\mathrm{NL}}^{\mathrm{local}}\simeq
 \frac{5a_B^4\beta M_p^4}{4\gamma^3\phi_B'^5}\frac{1}{\Delta\eta_{\mathrm{amp}}}
 \left(\frac{\phi_B'}{\phi'(\eta_-)}\right)^5~.
\end{equation}

In order to combine the bound on curvature perturbations
and the above result,
it is useful to rewrite the expressions
for $f_{\mathrm{NL}}$ in terms of the ratio
$\Delta\zeta/\zeta(\eta_{B-})$.
In Sec.\ \ref{sec:pert}, Eq.\ \eqref{eq:ubDeltazeta}
told us that
\begin{equation}\label{eq:Dzoz1}
 \frac{\Delta\zeta}{\zeta(\eta_{B-})}\lesssim 2\frac{\zeta'(\eta_{B-})}{\zeta(\eta_{B-})}\left(\frac{\phi'_B}{\phi'(\eta_{-})}\right)^2\Delta\eta_{\mathrm{amp}}~.
\end{equation}
In the previous section, we argued that the initial
conditions at $\eta_{B-}$ were given by the end conditions
of the matter-dominated phase of contraction, so we can say
that
\begin{equation}
 \zeta(\eta_{B-})\stackrel{k/\mathcal{H}\rightarrow 0}{\simeq}\zeta_k^{\mathrm{m}}(\eta_{B-})~,
 \qquad \zeta'(\eta_{B-})\stackrel{k/\mathcal{H}\rightarrow 0}{\simeq}\zeta_k^{\mathrm{m}\prime}(\eta_{B-})~.
\end{equation}
Recalling that $\zeta_k^{\mathrm{m}}$ is given by
Eq.\ \eqref{eq:zetam}, we find that
\begin{equation}
 \frac{\zeta'(\eta_{B-})}{\zeta(\eta_{B-})}
 \simeq\lim_{k/\mathcal{H}\rightarrow 0}\frac{\zeta_k^{\mathrm{m}\prime}(\eta_{B-})}{\zeta_k^{\mathrm{m}}(\eta_{B-})}
 =\frac{3}{\Delta\eta_{B-}}~,
\end{equation}
and thus Eq.\ \eqref{eq:Dzoz1} becomes
\begin{equation}
 \frac{1}{6}\left(\frac{\Delta\eta_{B-}}{\Delta\eta_{\mathrm{amp}}}\right)\left(\frac{\Delta\zeta}{\zeta(\eta_{B-})}\right)
 \lesssim\left(\frac{\phi'_B}{\phi'(\eta_{-})}\right)^2~.
\end{equation}
This allows us to place a lower bound on
Eq.\ \eqref{eq:fNLlocalsimpl1} as follows,
\begin{equation}\label{eq:fNLlocallb}
 f_{\mathrm{NL}}^{\mathrm{local}}\gtrsim\frac{5a_B^4\beta M_p^4}{144\sqrt{6}\gamma^3\phi_B'^5}\frac{1}{\Delta\eta_{\mathrm{amp}}}
 \left(\frac{\Delta\eta_{B-}}{\Delta\eta_{\mathrm{amp}}}\right)^{5/2}\left(\frac{\Delta\zeta}{\zeta(\eta_{B-})}\right)^{5/2}~.
\end{equation}

Our initial estimation in Sec.\ \ref{sec:bispectrumintro}
showed that we expected $f_{\mathrm{NL}}$
to have terms of order $(\Delta\zeta/\zeta)^1$ and
$(\Delta\zeta/\zeta)^2$. The terms of order
$(\Delta\zeta/\zeta)^1$ in the full calculation corresponded
to terms of order $[\phi_B'/\phi'(\eta_-)]^2$
in our approximation scheme and they originated
from $\mathcal{A}_{\zeta\zeta'^2}$
and $\mathcal{A}_{\zeta'^3}$.
A term of order $(\Delta\zeta/\zeta)^2$,
i.e.\ of order $[\phi_B'/\phi'(\eta_-)]^4$,
could have originated from $\mathcal{A}_{\zeta\zeta'^2}$
but the full calculation showed that it did not
have any real component [see Eq.\ \eqref{eq:AIM}].
Instead, the full calculation showed the presence
of terms of order $(\Delta\zeta/\zeta)^{5/2}$,
$(\Delta\zeta/\zeta)^{3/2}$,
and $(\Delta\zeta/\zeta)^{1/2}$
coming from the field redefinition
contribution to the shape function.
In the large amplification limit,
we are left with one term of order
$(\Delta\zeta/\zeta)^{5/2}$
as shown in Eq.\ \eqref{eq:fNLlocallb}.

Let us recall that $\Delta\zeta/\zeta$ is bounded
from below in order to satisfy the current observational
bound on the tensor-to-scalar ratio $r$.
Using the bound\ \eqref{eq:constraintfromr},
we can further constrain the bound\ \eqref{eq:fNLlocallb},
\begin{equation}\label{eq:fNLlocallb2}
 f_{\mathrm{NL}}^{\mathrm{local}}\gtrsim 240\left(\frac{\beta}{\gamma^3}\right)
 \left(\frac{a_BM_p}{\phi_B'}\right)^5
 \left(\frac{(a_BM_p)^{-1}}{\Delta\eta_{\mathrm{amp}}}\right)
 \left(\frac{\Delta\eta_{B-}}{\Delta\eta_{\mathrm{amp}}}\right)^{5/2}~.
\end{equation}
Let us note that $\Delta\eta_{B-}\sim\mathcal{H}_{B-}^{-1}$,
and since the bounce energy scale is taken to be
much less than the Planck scale, it results
that $\Delta\eta_{B-}\gg (a_BM_p)^{-1}$.
Thus, since every dimensionless ratio in
Eq.\ \eqref{eq:fNLlocallb2} at least of order 1
or much greater than 1, it results that
$f_{\mathrm{NL}}^{\mathrm{local}}\gtrsim 240$.
Including the negative contribution to
$f_{\mathrm{NL}}^{\mathrm{local}}$
from the matter-dominated contracting phase
which is of order 1\ \cite{Cai:2009fn}
and the negative contributions from
$\mathcal{A}_{\zeta\zeta'^2}$ and
$\mathcal{A}_{\zeta'^3}$
would reduce this bound,
but really not significantly.

The best observational bounds on primordial non-Gaussianities
currently come from the Planck experiment,
which reports\ \cite{Ade:2015ava}
\begin{equation}
 f_{\mathrm{NL}}^{\mathrm{local}}=0.8\pm5.0~,
 \qquad f_{\mathrm{NL}}^{\mathrm{equil}}=-4\pm43~,
 \qquad f_{\mathrm{NL}}^{\mathrm{ortho}}=-26\pm21~,
\end{equation}
at $68\%$ CL. We see that the lower bound
on $f_{\mathrm{NL}}^{\mathrm{local}}$ coming from
the bounce phase is already excluded by the observations
at very high confidence level. Following the same
analysis as above for the equilateral and
orthogonal shapes yields the bounds
$f_{\mathrm{NL}}^{\mathrm{equil}}\gtrsim 359$
and $f_{\mathrm{NL}}^{\mathrm{ortho}}\gtrsim 289$,
which are also excluded at very high confidence level,
although to a smaller extent than
$f_{\mathrm{NL}}^{\mathrm{local}}$.

To summarize, in this section we took the non-Gaussianity
results derived in the previous section and imposed that
there had been a sufficient amplification of
curvature perturbations in order to satisfy
the current observational bound on the
tensor-to-scalar ratio. As a result,
the theoretical lower bounds on
$f_{\mathrm{NL}}^{\mathrm{local}}$,
$f_{\mathrm{NL}}^{\mathrm{equil}}$,
and $f_{\mathrm{NL}}^{\mathrm{ortho}}$
are excluded at high confidence level
by the current observational constraints
on non-Gaussianities.
This shows that the model suffers from the
``no-go'' theorem that we conjectured in
Sec.\ \ref{sec:thenogoconjecture}.

Looking at Eq.\ \eqref{eq:fNLlocallb2}, we see that this
could be alleviated if, for instance, the amplification
period was very long compared to the Planck time,
or if the model parameters were such that
$\beta/\gamma^3\ll 1$ or $a_BM_p/\phi_B'\ll 1$.
However, these conditions seem unlikely to occur
in a physically admissible matter bounce scenario.

\section{Conclusions}\label{sec:conclusion}

In the present paper, we have studied in detail the
nonlinear dynamics of primordial curvature perturbations during the nonsingular
bouncing phase in a matter bounce model described by a single generic scalar
field minimally coupled to Einstein gravity. This type of model can be made
consistent with the observational bound on the tensor-to-scalar ratio by
realizing an enhancement of the curvature perturbations in the bouncing phase.
We derived the conditions on the model parameters for which such an
enhancement can be achieved. We then expanded the action for perturbations
up to the third order, computed the full set of three point correlation functions
and then derived the nonlinearity parameters $f_{\rm NL}$ in the cases of
specific shapes of observational interest. Our results show that if the primordial
non-Gaussianities mainly arise from a manifest growth of curvature perturbations
during the bounce, then the nonlinearity parameter would become dangerously large
and lead to a disagreement with the observational
constraints from cosmic microwave background (CMB) data\footnote{We recall
that it has also been found in\ \cite{Gao:2014eaa,Gao:2014hea}
that non-Gaussianities could become dangerously large in a certain
nonsingular bouncing cosmology and it has been conjectured that this
could be generic to a large family of nonsingular bouncing cosmologies.}.
Specifically, we examined the relation between the nonlinearity parameter
in the local, equilateral, and orthogonal limits
and the growth of the curvature perturbations and explicitly
showed that the observational bounds on the tensor-to-scalar ratio and
the CMB bispectrum cannot be simultaneously satisfied. This leads us to
conjecture that there is a ``no-go'' theorem for single field matter bounce cosmologies,
assuming that the nonsingular bounce is realized by a generic scalar field
minimally coupled to Einstein gravity, {\it{which would rule out a large class of
matter bounce models}}.

We note that this ``no-go" theorem might be circumvented by dropping
certain assumptions imposed above. For instance, if one introduces more than one degree
of freedom such as in the matter bounce curvaton mechanism\ \cite{Cai:2011zx,Cai:2014bea}, the
constraints from the tensor-to-scalar ratio as well as from the primordial non-Gaussianities
can be satisfied at the same time, the reason being that in the curvaton scenario the
scalar fluctuations are generated by a different mechanism than the tensors.
As another example, if the initial Bunch-Davies vacuum is noncanonical
(e.g., in the $\Lambda$CDM bounce\ \cite{Cai:2014jla}, the initial quantum vacuum has $c_s\ll 1$),
the initial ratio of the tensor modes to the scalar modes can be suppressed, in which case
there is no need for the curvature perturbations to be enhanced during the bounce.

Our analysis also does not immediately apply to nonsingular
matter bounce models in which the violation of the null energy condition
is obtained by changes in the gravitational action
(e.g., in Loop Quantum Cosmology\ \cite{WilsonEwing:2012pu,Cai:2014zga},
Ho\v{r}ava-Lifshitz gravity\ \cite{Brandenberger:2009yt},
extended $F(R)$ gravity\ \cite{Bamba:2013fha, Nojiri:2014zqa},
modified Gauss-Bonnet gravity\ \cite{Bamba:2014mya}, or torsion gravity
scenarios\ \cite{Cai:2011tc, Amoros:2013nxa}).
We might expect that the no-go theorem will extend to modified gravity
matter bounce models which have the same number of degrees of freedom as
General Relativity, in which case the tensor-to-scalar is generically
large\ \cite{Cai:2014xxa}. However, if the gravity model contains new
degrees of freedom, then we might be in a situation similar to what happens
in the curvaton scenario: the new degrees of freedom source the scalar
modes but not the tensor modes, thus suppressing the tensor-to-scalar ratio
during the bounce phase. Yet, it would be interesting to explicitly analyze
the conditions under which the bispectrum constraints can be made consistent
with the observed bound on the tensor-to-scalar ratio in such models.

\section*{Acknowledgments}

We thank Keshav Dasgupta, Evan McDonough, and Yi Wang for valuable discussions.
J.Q.~acknowledges the Fonds de recherche du Qu\'{e}bec - Nature et technologies (FRQNT) and
the Walter C. Sumner Foundation for financial support. The research at McGill is
supported in part by an NSERC Discovery grant (R.B.) and by funds from
the Canada Research Chair program (R.B.). Y.F.C. is supported in part by the Chinese
National Youth Thousand Talents Program and in part by the USTC
start-up funding under Grant No. KY2030000049.

\appendix

\section{CURVATURE PERTURBATIONS EXPANDING ABOUT THE SINGULARITY}\label{sec:curv_pert_sing}
The equation of motion for curvature perturbations in the IR limit is
[see Eq.\ \eqref{EOMR}]
\begin{equation}
 \frac{d\zeta'}{d\eta}+\frac{(z^2)'}{z^2}\zeta'=0~.
\end{equation}
Let us parametrize $z^2$ close to the singular point $\eta_s$ as
\begin{equation}
 z^2(\eta)\sim\frac{1}{(\eta-\eta_s)^2}~,
\end{equation}
so the equation of motion becomes
\begin{equation}
 \frac{d\zeta'}{d\eta}=\frac{2}{\eta-\eta_s}\zeta'~.
\end{equation}
Since after the singular time we have $\eta>\eta_s>\eta_i$,
we integrate as follows,
\begin{equation}
 \ln\left(\frac{\zeta'(\eta)}{\zeta'(\eta_i)}\right)=2\int_{\eta_i}^\eta\frac{d\tilde\eta}{\tilde\eta-\eta_s}=2\left(\int_{\eta_i}^{\eta_s-\epsilon}+\int_{\eta_s-\epsilon}^{\eta_s+\epsilon}+\int_{\eta_s+\epsilon}^{\eta}\right)\frac{d\tilde\eta}{\tilde\eta-\eta_s}~,
\end{equation}
for some constant $\epsilon$. As we take the limit $\epsilon\rightarrow 0$,
the second integral vanishes by definition and we are left with
the first and third integral. Evaluating them, we find
\begin{align}
 \ln\left(\frac{\zeta'(\eta)}{\zeta'(\eta_i)}\right)=&~2\lim_{\epsilon\rightarrow0}\left[\ln\left(\frac{(\eta_s-\epsilon)-\eta_s}{\eta_i-\eta_s}\right)+\ln\left(\frac{\eta-\eta_s}{(\eta_s+\epsilon)-\eta_s}\right)\right] \nonumber \\
 =&~2\lim_{\epsilon\rightarrow0}\ln\left(\frac{-\epsilon(\eta-\eta_s)}{(\eta_i-\eta_s)\epsilon}\right) \nonumber \\
 =&~2\ln\left(\frac{\eta-\eta_s}{\eta_s-\eta_i}\right)~.
\end{align}
Therefore,
\begin{equation}
 \zeta'(\eta)=\zeta'(\eta_i)\left(\frac{\eta-\eta_s}{\eta_s-\eta_i}\right)^2
\end{equation}
as expected if there were no singularity. A final integration yields
\begin{equation}
 \zeta(\eta)=\zeta(\eta_i)+\zeta'(\eta_i)\left(\frac{(\eta-\eta_s)^3+(\eta_s-\eta_i)^3}{3(\eta_s-\eta_i)^2}\right)~.
\end{equation}
As expected, we recover the constant mode solution $\zeta'=0$ as $\eta\rightarrow\eta_s$.

\section{PERTURBATIONS OUTSIDE THE BOUNCE PHASE}\label{sec:pertoutsidebounce}

Let us consider matter with an equation of state $P=w\rho$ with $w$
independent of time. In this case $z(t)\sim a(t) M_p$
(see Sec.\ \ref{sec:flucmatterbounce}). Then, the solution
to the long wavelength curvature perturbations is given by
[see Eq.\ \eqref{eq:dotzetasol}]
\begin{align}
 \zeta(t)&=\zeta(t_i)+\dot\zeta(t_i)a(t_i)z(t_i)^2\int_{t_i}^{t}\frac{d\tilde t}{a(\tilde t)z^2(\tilde t)} \\
 &=\zeta(t_i)+\dot\zeta(t_i)a(t_i)^3\int_{t_i}^{t}\frac{d\tilde t}{a^3(\tilde t)}~.
\end{align}
For a constant $w\neq -1$, the solution to the scale factor is given
by
\begin{equation}
 a(t) \, = \, a_0t^{2/3(1+w)}~,
\end{equation}
for some positive constant $a_0$, so we find
\begin{align}
 \zeta(t)&=\zeta(t_i)+\dot\zeta(t_i)t_i^{\frac{2}{1+w}}\int_{t_i}^{t}d\tilde t~\tilde t^{-\frac{2}{1+w}} \\
 &=\zeta(t_i)+\dot\zeta(t_i)t_i^{\frac{2}{1+w}}\left(\frac{w+1}{w-1}\right)\left(t^{\frac{w-1}{w+1}}-t_i^{\frac{w-1}{w+1}}\right)~,
\end{align}
as long as $|w|\neq 1$.
Thus, for matter with $|w|>1$, the solution for $\zeta$ exhibits
a constant mode and a mode which is growing in an expanding universe
(decaying in a contracting background), whereas for matter with
$|w|<1$, it exhibits a constant mode and a mode which is decaying in
an expanding universe (and growing in a contracting background).
For example, this implies that an Ekpyrotic phase of contraction
in which $w\gg 1$ has a constant mode and a decaying mode.

For $w=-1$, one would recover the standard result of inflation where
the constant mode is dominant on super-Hubble scales in an expanding
background, and the second mode dominates in a contracting space.

The $w=1$ case of fast roll expansion is relevant for the dynamics of
our nonsingular bouncing cosmology right after the bounce phase.
A phase of fast roll expansion occurs if the Lagrangian for
the scalar field is dominated by its kinetic term,
i.e.\ $V(\phi)\ll\dot\phi^2/2$. It then follows that the solution for
the curvature perturbations in this case is (here in conformal time)
\begin{equation}
 \zeta(\eta) \,
 = \, \zeta(\eta_i)+\zeta'(\eta_i)a(\eta_i)^2\int_{\eta_i}^{\eta}\frac{d\tilde\eta}{a^2(\tilde\eta)}~.
\end{equation}
Solving the background dynamics tells us that the solution to the
scale factor in a phase of fast roll expansion is
\begin{equation}
 a(\eta) \, = \, c_E(\eta-\eta_E)^{1/2}~,
\end{equation}
where $c_E$ and $\eta_E$ are some constants. Thus,
\begin{align}
 \zeta(\eta)&=\zeta(\eta_i)+\zeta'(\eta_i)(\eta_i-\eta_E)\int_{\eta_i}^{\eta}\frac{d\tilde\eta}{\eta-\eta_E} \\
 &=\zeta(\eta_i)+\zeta'(\eta_i)(\eta_i-\eta_E)\ln\left(\frac{\eta-\eta_E}{\eta_i-\eta_E}\right)~.
\end{align}
So, for $w=1$, curvature perturbations exhibit a constant mode solution
and a logarithmically growing mode, i.e.\ $\zeta(\eta)\sim\ln\eta$.
We note that this is also true in physical time since $a\sim t^{1/3}\sim\eta^{1/2}$
implies that $\zeta(t)\sim\ln t^{2/3}\sim\ln t$.

\section{THIRD ORDER PERTURBED ACTION}

\subsection{Derivation of the general form of the third order action}\label{sec:thirdorderaction}

To study the three point correlation function in this matter bounce model,
we have to evaluate the action up to third order in
perturbation theory. We use the metric in the Arnowitt-Deser-Misner (ADM) form (see, e.g.,\ \cite{Wang:2013zva})
\begin{equation}\label{eq:metric}
ds^2=N^2 dt^2-\gamma_{ij} (N^i dt+dx^i)(N^j dt+dx^j)~,
\end{equation}
where $N_i=\gamma_{ij} N^j$ is the shift vector and $N$ is the lapse function.
The tensor $\gamma_{ij}$ is the metric of the 3-dimensional spacelike hypersurfaces
in this $3+1$ decomposition.
It is related to the full 4-dimensional space-time metric tensor $g_{\mu\nu}$ via
$\sqrt{-g}=N\sqrt{\gamma}$,
where $g$ and $\gamma$ are the determinants of the tensors $g_{\mu\nu}$ and $\gamma_{ij}$,
respectively.
The action\ \eqref{actionunperturbed} in this ADM decomposition is given by
\begin{equation}\label{eq:action}
S=\int d^4x ~ \sqrt{-g}\left[\frac{M_p ^2}{2}\left(R^{(3)} +\kappa_{ij} \kappa^{ij} -\kappa ^2\right)+K(\phi,X)+G(\phi,X)  \square\phi\right]~,
\end{equation}
where $R^{(3)}$ is the three-dimensional Ricci scalar and the extrinsic curvature is defined by
\begin{equation}\label{eq:ext_curv}
 \kappa_{ij}\equiv\frac{1}{2 N} \left(\dot{\gamma}_{ij}-D_i N_i-D_j N_i\right)~.
\end{equation}
We define the covariant derivative $D_i$ on the spacelike hypersurfaces such that
it is torsion-free and satisfies
\begin{equation}
 D_i\gamma^{ij}=0~.
\end{equation}
Then, $R^{i(3)}_{\ jkl}$ is the Riemann tensor associated with this connection, and
\begin{align}
 R^{(3)}_{ij}&=R^{k(3)}_{\ ikj}~, \\
 R^{(3)}&=\gamma^{ij}R^{(3)}_{ij}~,
\end{align}
are the Ricci tensor and Ricci scalar, respectively.
In the uniform field gauge where
\begin{align}
 \delta\phi&=0~, \\
 \gamma_{ij}&=a^2 e^{2 \zeta} \delta _{ij}~,
\end{align}
one can use the Hamiltonian and momentum constraints to determine the scalar
contributions to the lapse function and shift vector,
\begin{equation}\label{eq:pert_var}
 N=1+\alpha~, \ N_i=\partial_i \sigma~,
\end{equation}
up to leading order.
Substituting Eq.\ \eqref{eq:pert_var} into the metric [Eq.\ \eqref{eq:metric}]
and expanding the action [Eq.\ \eqref{eq:action}]
up to third order, we obtain the following,
\begin{align}\label{eq:S3}
S_3=&\int d^4x ~ a^3 \left[a_1 \alpha ^3+a_2 \zeta \alpha^2 +a_3 \dot{\zeta} \alpha^2+a_4 \frac{\partial^2 \sigma}{a^2} \alpha^2+a_5 \frac{\partial \zeta \partial \sigma}{a^2} \alpha+a_6 \alpha \dot{\zeta} \zeta+ a_7 \alpha \zeta \frac {\partial ^2 \sigma}{a^2} +3 M_p ^2 \alpha \dot{\zeta}^2\right. \nonumber \\
&-\frac{M_p ^2}{2 } \ \frac{(\partial_i \partial_j \sigma)^2-(\partial^2 \sigma)^2}{a^4} \alpha+2 M_p ^2 H \zeta \frac{\partial \zeta \partial \sigma}{a^2} -2 M_p ^2 \dot{\zeta} \alpha \frac{\partial^2 \sigma}{a^2}-2 M_p ^2 \zeta \alpha \frac{\partial^2 \zeta}{a^2}-M_p ^2 \zeta^2 \frac{\partial^2 \zeta}{a^2}-M_p ^2 \alpha \frac{(\partial \zeta)^2}{a^2} \nonumber \\
&-M_p ^2 \zeta \frac{(\partial \zeta)^2}{a^2}-9 M_p ^2 \dot{\zeta}^2 \zeta+2 M_p ^2 \dot{\zeta} \frac{\partial \zeta \partial \sigma}{a^2}+M_p ^2 H \zeta^2 \frac{\partial^2 \sigma}{a^2}+2 M_p ^2 \dot{\zeta} \zeta \frac{\partial^2\sigma}{a^2}-\frac{M_p ^2}{2 } \ \frac{(\partial_i \partial_j \sigma)^2-(\partial^2 \sigma)^2}{a^4} \zeta \nonumber \\
&\left.-2 M_p ^2 \frac{\partial_i \zeta \partial_j \sigma \partial_i \partial_j \sigma}{a^4}\right]~,
\end{align}
where we defined the following coefficients,
\begin{align}
a_1\equiv&~3 M_p^2 H^2 - \dot{\phi}^2 \left( \frac{1}{2} K_{,X} + \dot{\phi}^2 K_{,XX} + \frac{1}{6} \dot{\phi}^4 K_{,XXX} \right) - 2 H \dot{\phi}^3 \left(5 G_{,X} + \frac{11}{4} \dot{\phi}^2 G_{,XX} + \frac{1}{4} \dot{\phi}^4 G_{,XXX}\right) \nonumber \\
&+ \dot{\phi}^2 \left( G_{,\phi} + \frac{7}{6} \dot{\phi}^2 G_{,X \phi} + \frac{1}{6} \dot{\phi}^4 G_{,\phi XX}\right)~, \nonumber \\
a_2\equiv&-9 M_p^2 H^2 + 3 \dot{\phi}^2 \left( \frac{1}{2} K_{,X} + \frac{1}{2} \dot{\phi}^2 K_{,XX} \right) + 18 H \dot{\phi}^3 \left( G_{,X} + \frac{1}{4} \dot{\phi}^2 G_{,XX} \right) - 3 \dot{\phi}^2 \left( G_{,\phi} + \frac{1}{2} \dot{\phi}^2 G_{,\phi X} \right)~, \nonumber \\
a_3 \equiv&-6 M_p^2 H + 6 \dot{\phi}^3 \left( G_{,X}+\frac{1}{4} \dot{\phi}^2 G_{,XX} \right)~, \nonumber \\
a_4 \equiv&~2 M_p^2 H - 2 \dot{\phi}^3 \left( G_{,X} + \frac{1}{4} \dot{\phi}^2 G_{,XX} \right)~, \nonumber \\
a_5 \equiv&-2 M_p^2 H+3 \dot{\phi}^3 G_{,X}~, \nonumber \\
a_6 \equiv&-9 \left( -2 M_p^2 H + \dot{\phi}^3 G_{,X} \right)~, \nonumber \\
a_7 \equiv&-2 M_p^2 H + 3 \dot{\phi}^3 G_{,X}~. \nonumber
\end{align}
We note that the Hamiltonian and momentum constraints yield (these can also be obtained
by varying the action above with respect
to $\alpha$ and $\sigma$)
\begin{align}\label{eq:constraint1}
\alpha =&~ \frac{2 M_p ^2 \dot{\zeta}}{u}~, \\
\label{eq:constraint2}
\partial^2 \sigma =&~ a_8 \partial^2 \zeta + \partial^2 \chi~,
\end{align}
respectively, where we defined
\begin{align}
 u&\equiv 2 M_p ^2 H- \dot{\phi}^3 G_{,X}~, \\
 a_8&\equiv -\frac{2 M_p^2}{u}~,
\end{align}
and where
\begin{equation}\label{eq:d2chi}
 \partial^2\chi\equiv\frac{z^2\dot{\zeta}}{M_p ^2}= 3a^2\dot{\zeta}+\frac{2 M_p^2 a^2 \dot{\zeta}}{u^2} \left(-6 M_p^2 H^2+\dot{\phi}^2 K_{,X}+ \dot{\phi}^4 K_{,XX}+12 H \dot{\phi}^3 G_{,X}+3 H \dot{\phi}^5 G_{,XX}-2 \dot{\phi}^2 G_{,\phi}-\dot{\phi}^4 G_{,X \phi}\right)~.
\end{equation}
If we substitute Eqs.\ \eqref{eq:constraint1} and\ \eqref{eq:constraint2} into
the third order perturbed action [Eq.\ \eqref{eq:S3}], we obtain
\begin{align}\label{eq:S3_2}
S_3=&\int d^4x ~ a^3 \left[ A_1 \dot{\zeta}^3+ A_2 \zeta \dot{\zeta}^2+A_3 (\partial ^2 \zeta) \dot{\zeta}^2 +A_4 \dot{\zeta} (\partial \zeta)^2 +A_5 \partial \zeta \partial \chi \dot{\zeta}+A_6 \zeta \dot{\zeta} \partial ^2 \zeta+A_7 \zeta (\partial \zeta)^2+A_8 \partial \zeta \partial \chi \zeta \right. \nonumber \\
&+ A_9 \dot{\zeta}^2 \partial ^2 \chi+A_{10} \zeta ^2 \partial ^2 \zeta+A_{11} \zeta ^2 \partial ^2 \chi +A_{12} \partial _i \zeta \partial _j \zeta \partial_i \partial_j \zeta+A_{13} \partial _i \zeta \partial _j \zeta \partial_i \partial_j \chi+ A_{14} \partial _i \zeta \partial _j \chi \partial_i \partial_j \zeta \nonumber \\
&+ A_{15} \partial _i \zeta \partial _j \chi \partial_i \partial_j \chi+(A_{16} \dot{\zeta} +A_{17} \zeta) (\partial _i \partial _j \zeta)^2+(A_{18} \dot{\zeta} +A_{19} \zeta) (\partial _i \partial _j \chi)^2+ (A_{20} \dot{\zeta} +A_{21} \zeta) \partial _i \partial _j \zeta \partial _i \partial _j \chi \nonumber \\
&+ \left.(A_{22} \dot{\zeta} +A_{23} \zeta) (\partial ^2 \zeta)^2+(A_{24} \dot{\zeta} +A_{25} \zeta) (\partial ^2 \chi)^2+ (A_{26} \dot{\zeta} +A_{27} \zeta) \partial ^2 \zeta \partial ^2 \chi+ A_{28} \zeta \dot{\zeta} \partial ^2 \chi \right]+S_b~,
\end{align}
where we defined the following,
\begin{eqnarray}
&&A_1 \equiv \frac{(2 M_p^2)^3}{u^3} a_1 + \frac{(2 M_p^2)^2}{u^2} a_3 +6 \frac{M_p^4}{u}~,\ \ A_2 \equiv\frac{(2 M_p^2)^2}{u^2} a_2 + \frac{2 M_p^2}{u} a_6 -9 M_p ^2~, \nonumber \\
&&A_3 \equiv \frac{(2 M_p^2)^2}{u^2 a^2} a_4 a_8 - \frac{(2 M_p^2)^2}{u a^2} a_8~,\ \ A_4 \equiv \frac{2 M_p^2}{u a^2} a_5 a_8 - \frac{2 M_p^4}{u a^2} +\frac{2 M_p^2}{a^2} a_8~, \nonumber \\
&&A_5 \equiv \frac{2 M_p^2}{u a^2} a_5 +\frac{2 M_p^2}{a^2}~,\ \ A_6 \equiv \frac{2 M_p^2}{u a^2} a_8 a_7 -\frac{(2 M_p^2)^2}{u a^2} +\frac{2 M_p^2}{a^2} a_8~, \nonumber \\
&&A_7 \equiv \frac{2 M_p^2}{a^2} H a_8 -\frac{M_p^2}{a^2}~,\ \ A_8 \equiv \frac{2 M_p^2 H}{a^2}~,\ \ A_9 \equiv \frac{(2 M_p^2)^2}{u^2 a^2} a_4 - \frac{(2 M_p^2)^2}{u a^2}~, \nonumber \\
&&A_{10} \equiv -\frac{M_p^2}{a^2} +\frac{ M_p^2}{a^2} H a_8~,\ \ A_{11}\equiv\frac{M_p^2 H}{a^2}~,\ \ \frac{A_{12}}{a_8}\equiv A_{13}\equiv A_{14}\equiv-\frac{2 M_p^2}{a^4}  a_8~, \nonumber \\
&&A_{15} \equiv -\frac{2 M_p^2}{a^4}~,\ \ A_{16}\equiv-\frac{M_p^4}{u a^4} a_8 ^2~,\ \ A_{17}\equiv-\frac{M_p^2}{2 a^4}  a_8 ^2~,\ \ A_{18} \equiv -\frac{M_p^4}{u a^4}~, \nonumber \\
&&A_{19} \equiv -\frac{M_p^2}{2 a^4}~,\ \ A_{20}\equiv-\frac{2 M_p^4}{u a^4}  a_8~,\ \ A_{21}\equiv-\frac{M_p^2}{a^4}  a_8~,\ \ A_{22}\equiv\frac{M_p^4}{u a^4}  a_8 ^2~, \nonumber \\
&&A_{23} \equiv \frac{M_p^2}{2 a^4}  a_8 ^2~,\ \ A_{24}\equiv\frac{M_p^4}{u a^4}~,\ \ A_{25}\equiv\frac{M_p^2}{2 a^4}~,\ \ A_{26} \equiv \frac{2 M_p^4}{u a^4} a_8~, \nonumber \\
&&A_{27} \equiv \frac{M_p^2}{a^4} a_8~,\ \ A_{28}\equiv\frac{2 M_p^2}{a^2}+\frac{2 M_p^2}{u a} a_7~. \nonumber
\end{eqnarray}
We note that $S_b$ is a short-hand notation for all the boundary terms, which do not
make a contribution to the calculation at 3rd order.
After many integrations by part, we obtain
\begin{align}\label{eq:S3_3_appendix}
S_3=&\int d^4x~ \left( B_1 \left[\partial \zeta \partial \chi \partial ^2 \zeta- \zeta \partial _i \partial _j (\partial _i \zeta \partial_j \chi) \right]+ B_2 \dot{\zeta}^2 \partial ^2 \zeta \right. \nonumber \\
&+\left.B_3 \dot{\zeta} \partial \zeta \partial{\chi} +B_4 \zeta (\partial_i \partial_j\chi)^2 + B_5 \zeta (\partial \zeta)^2 +B_6 \dot{\zeta} ^3 +B_7 \dot{\zeta} ^2 \zeta-2f(\zeta) \frac{\delta {\cal{L}}_2}{\delta \zeta} \right)~,
\end{align}
where
\begin{equation}
 \frac{\delta {\cal{L}}_2}{\delta \zeta}= \partial _t(a z^2 \dot{\zeta})-\frac{c_s^2 z^2}{a} \partial ^2 \zeta~,
\end{equation}
and where
\begin{equation}\label{eq:f_appendix}
f(\zeta)=\frac{A_{20} a^2}{4 M_p^2} \left[ (\partial \zeta)^2 -\partial ^{-2} \partial _i  \partial_ j (\partial _i \zeta \partial_j \zeta) \right] + \frac{A_{18} a^2}{M_p^2} \left[ \partial \zeta \partial \chi - \partial ^{-2} \partial _i \partial _j (\partial _i \zeta \partial_j \chi) \right] - \frac{2 A_4 a^3 - C_1}{2 z^2 c_s^2} a \zeta \dot{\zeta}~.
\end{equation}
We also introduced the following,
\begin{align}\label{eq:B_3_appendix}
B_1\equiv&-A_{21} a ^3-\frac{1}{2} \partial _t (A_{20} a^3)+\frac{A_{20}}{2} a^3 H- 2 A_{18}z^2 c_s ^2 a~, \nonumber \\
B_2\equiv&~A_3 a^3 +(A_{26}+A_{20}) \frac{z^2a^3}{M_p^2}~, \nonumber \\
B_3\equiv&~A_5 a^3 +A_{15} \frac{z^2a^3}{M_p^2}~, \nonumber \\
B_4\equiv&-\partial _t (A_{18}a^3)-3A_{19}a^3+2A_{18} H a^3~, \nonumber \\
B_5\equiv&~\partial _t \left(A_{4}+A_{13}\frac{z^2}{2M_p^2}\right)a^3-2A_{10}a^3+A_{7}a^3~, \nonumber \\
B_6\equiv&~A_1 a^3 + (A_{18} +A_{24}) \left(\frac{z^2}{M_p^2} \right)^2 a^3 +A_9 \frac{z^2a^3}{M_p^2}+\frac{2 A _4 a^3 -C_1}{2 c_s^2} a^2~, \nonumber \\
B_7\equiv&~A_2 a^3+\left[A_{15} a^3 + \partial_ t (A_{18} a) a^2-  \partial_ t (A_{18} a^3) -3 A_{19} a^3 +2 A_{18} H a^3 + A_{25} a^3\right]\left(\frac{z^2}{M_p^2}\right)^2 \nonumber \\
&+ \frac{a z^2}{2}  \partial_ t \left[\frac{ (2 A_{4} a^3- C_1) a}{c_s ^2 z^2}\right] - \partial_ t \left(\frac{a z^2}{2}\right)  \frac{ (2 A_{4} a^3- C_1) a}{c_s ^2 z^2}-B_4 \left(\frac{z^2}{M_p^2}\right)^2~.
\end{align}
Furthermore,
\begin{equation}
 C_1\equiv A_6 a^3+ (A_{27}-A_{21})\frac{z^2a^3}{M_p^2}~,
\end{equation}
and $c_s$ is the speed of sound introduced earlier in Eq.\ \eqref{cs2}.

\subsection{Third order perturbed action in the limit of the matter-dominated contracting phase}\label{sec:S3limitmatter}

Let us evaluate the third order action given by Eq.\ \eqref{eq:S3_3_appendix}
in a matter-dominated contracting phase when
$G(\phi,X)=0$ and $K(\phi,X)=M_p^2 X-V(\phi)$. In this case, we have
\begin{eqnarray}
&&a_1=3 M_p^2 H^2 - \frac{1}{2}M_p^2 \dot{\phi}^2~,\ \ a_2=-9 M_p^2 H^2 + \frac{3}{2}M_p^2 \dot{\phi}^2~,\ \ a_3 =-6 M_p^2 H~, \nonumber \\
&&a_4=2 M_p^2 H~,\ \ a_5 =-2 M_p^2 H~,\ \ a_6 =18 M_p^2 H~,\ \ a_7 =-2 M_p ^2 H~, \nonumber
\end{eqnarray}
together with $u=2 M_p^2 H$, $a_8=-2 M_p^2/u$, and
$\partial ^2 \chi=z^2 \dot{\zeta}/M_p ^2=a^2 \dot{\phi}^2 \dot{\zeta}/2H^2$.
Then,
\begin{eqnarray}
&&A_1 = -\frac{M_p^2 \dot{\phi}^2}{2 H^3}~,\ \ A_2 =\frac{3 M_p^2 \dot{\phi}^2}{2 H^2}~,\ \ A_3 = 0~,\ \ A_4 = -\frac{M_p^2}{a^2 H}~,\ \ A_5 = 0~,\ \ A_6 = \-\frac{2 M_p^2}{a^2 H}~, \nonumber \\
&&A_7 = -\frac{3 M_p^2}{a^2}~,\ \ A_8 = \frac{2 M_p^2 H}{a^2}~,\ \ A_9 =0~,\ \ A_{10} = -\frac{2 M_p^2}{a^2}~,\ \ A_{11}=\frac{M_p^2 H}{a^2}~,\ \ \frac{A_{12}}{a_8}=A_{13}=A_{14}=\frac{2 M_p^2}{a^4 H}~, \nonumber \\
&&A_{15} = -\frac{2 M_p^2}{a^4}~,\ \ A_{16}=-\frac{M_p^4}{h a^4} a_8 ^2~,\ \ A_{17}=-\frac{M_p^2}{2 a^4} a_8^2~,\ \ A_{18} = -\frac{M_p^2}{2 a^4 H}~,\ \ A_{19}=-\frac{M_p^2}{2 a^4}~, \nonumber \\
&& A_{20}=-\frac{2 M_p^4}{h a^4} a_8~,\ \ A_{21}=-\frac{M_p^2}{a^4} a_8~,\ \ A_{22}=\frac{M_p^4}{h a^4} a_8^2~,\ \ A_{23} = \frac{M_p^2}{2 a^4} a_8^2~,\ \ A_{24}=\frac{M_p^4}{h a^4}~,\ \ A_{25}=\frac{M_p^2}{2 a^4}~, \nonumber \\
&&A_{26} = \frac{2 M_p^4}{h a^4} a_8~,\ \ A_{27} = \frac{M_p^2}{a^4} a_8~,\ \ A_{28}=\frac{2 M_p^2}{a^2}+\frac{2 M_p^2}{h a} a_7~. \nonumber
\end{eqnarray}
Thus,
\begin{align}
B_1(t)=&~M_p^2\frac{\dot{\phi}(t)^2+2 \dot{H}(t)}{2 a(t) H(t)^3}=0~, \nonumber \\
B_2(t)=&~0~, \nonumber \\
B_3(t)=&-\frac{M_p^2 a(t) \dot{\phi}(t)^2}{H(t)^2}=-2 \epsilon(t)  M_p^2 a(t)~, \nonumber \\
B_4(t)=&-\frac{M_p^2 \dot{H}(t)}{2 a(t) H(t)^2}=\frac{M_p^2 \epsilon(t) }{2 a(t)}~, \nonumber \\
B_5(t)=&~M_p^2 a(t) \frac{2 H(t)^2 \dot{H}(t)+2 \dot{\phi}(t) H(t) \ddot{\phi}(t)-\dot{\phi}(t)^2 \left[3 \dot{H}(t)-H(t)^2\right]}{2  H(t)^4}=M_p^2 \epsilon(t) ^2 a(t)~, \nonumber \\
B_6(t)=&~0~, \nonumber \\
B_7(t)=&~\frac{M_p^2 a(t)^3 \dot{\phi}(t) \left\{\dot{\phi}(t) \left[\dot{\phi}(t)^2+4 H(t)^2\right] \dot{H}(t)-8 H(t)^3 \ddot{\phi}(t)\right\}}{8 H(t)^6}=-\frac{1}{2} M_p^2 [\epsilon^3(t) -2 \epsilon^2(t)] a(t)^3~. \nonumber
\end{align}
Here, we consider $\dot{\phi}(t)^2=2 \epsilon(t) H(t)^2$, $\epsilon(t)=-\dot{H}(t)/H(t)^2$,
and $c_s=1$. Therefore, we find that
\begin{equation}\label{eq:S3_mb}
 S_3=\int d^4x~\left[-2 \epsilon  M_p^2 a(t) \partial \zeta \dot{\zeta} \partial{\chi} +\frac{M_p^2 \epsilon }{2 a(t)} \zeta (\partial_i \partial_j\chi)^2 + M_p^2 \epsilon ^2 a(t) \zeta (\partial \zeta)^2-\frac{1}{2} M_p^2 (\epsilon^3 -2 \epsilon ^2) a(t)^3 \dot{\zeta} ^2 \zeta-2 f(\zeta) \frac{\delta {\cal{L}}}{\delta \zeta}\right]~,
\end{equation}
and
\begin{equation}
 f(\zeta)=\frac{1}{4 a(t)^2 H(t)^2} \left[(\partial \zeta)^2 -\partial ^{-2} \partial _i  \partial_ j (\partial _i \zeta \partial_j \zeta)\right]+\frac{1}{2 a(t)^2 H(t)} \left[-\partial \zeta \partial \chi + \partial ^{-2} \partial _i \partial _j (\partial _i \zeta \partial_j \chi)\right] - \frac{1}{H(t)} \zeta \dot{\zeta}~.
\end{equation}
This is equivalent to the third order action given in\ \cite{Cai:2009fn}
noting that we defined $\partial^2 \chi=a^2 \epsilon \dot{\zeta}$
whereas they considered $\partial^2 \chi=\dot{\zeta}$.

\section{EVALUATING THE SHAPE FUNCTION IN THE BOUNCE PHASE}\label{sec:shape_function}

We want to evaluate Eq.\ \eqref{shapefunctionunevaluated}
which, as explained in the text, has three dominant terms:
the $\zeta\zeta'^2$ term, the $\zeta'^3$ term, and the
field redefinition term. Let us start with the contribution
from the $\zeta\zeta'^2$ term to the shape function,
which is
\begin{align}
    \mathcal{A}_{\zeta\zeta'^2}=&~\frac{\zeta_{k_3}^*(\eta_{+})}{4 \zeta_{k_1} (\eta_{+}) \zeta_{k_2} (\eta_{+})} \nonumber \\
      &\times i\int_{\eta_{-}}^{\eta_{+}}d\eta\left[\left(\frac{B_3(\eta)z^2(\eta)}{M_p^2a_B}k_3^3\frac{\vec{k}_1\cdot\vec{k}_3}{k_3^2}+\frac{B_4(\eta)z^4(\eta)}{M_p^4a_B}k_3^3\frac{(\vec{k}_2\cdot\vec{k}_3)^2}{k_3^2k_2^2}+\frac{B_7(\eta)}{a_B}k_3^3\right)\zeta_{k_1}(\eta)\zeta_{k_2}'(\eta)\zeta_{k_3}'(\eta)\right]~,
\end{align}
where we omit the 5 additional permutations for now.
Using Eq.\ \eqref{eq:zetabounceNL} for $\zeta_{k_i}(\eta)$, we get
\begin{align}\label{eq:Azzp2prelim1}
 \mathcal{A}_{\zeta\zeta'^2}=&~\frac{ik_3^3}{4a_B}\frac{\zeta_{k_3}^*(\eta_+)\zeta_{k_2}^{\mathrm{m}\prime}(\eta_{B-})\zeta_{k_3}^{\mathrm{m}\prime}(\eta_{B-})}{\zeta_{k_1}(\eta_+)\zeta_{k_2}(\eta_+)}\left(\frac{\phi_B'}{\phi'(\eta_-)}\right)^4 \nonumber \\
 &\times\int_{\eta_-}^{\eta_+}d\eta\left[\left(\frac{B_3(\eta)z^2(\eta)}{M_p^2}\frac{\vec{k}_1\cdot\vec{k}_3}{k_3^2}+\frac{B_4(\eta)z^4(\eta)}{M_p^4}\frac{(\vec{k}_2\cdot\vec{k}_3)^2}{k_3^2k_2^2}+B_7(\eta)\right)\zeta_{k_1}(\eta)\right]~.
\end{align}
We note that taking $\zeta_{k_i}'$ from Eq.\ \eqref{eq:zetabounceNL}
actually gives an upper bound on $\mathcal{A}_{\zeta\zeta'^2}$
since Eq.\ \eqref{eq:zetabounceNL} used the maximal
growth rate\ \eqref{eq:zetadotmax} for the full
range $[\eta_-,\eta_+]$. This introduces some small error
in the final result but this will turn out to be unimportant
since, as we will see, the field redefinition contribution
to the shape function will dominate over this upper bound on
$\mathcal{A}_{\zeta\zeta'^2}$.

The time-dependent terms that remain inside the integral
are $B_3$, $B_4$, $B_7$, $z^2$, and $\zeta_{k_1}$.
The latter, $\zeta_{k_1}$, may experience a nontrivial
enhancement during the interval $[\eta_-,\eta_+]$, and
consequently, it may contribute significantly to the integral.
The other terms, i.e.\ $B_3z^2$, $B_4z^4$, and $B_7$
defined in Eqs.\ \eqref{eq:B_3_appendix} and\ \eqref{z2_close_bounce},
contribute as almost constant terms in the integral over the range
$[\eta_-,\eta_+]$. Therefore, we rewrite Eq.\ \eqref{eq:Azzp2prelim1}
in the following form,
\begin{align}
 \mathcal{A}_{\zeta\zeta'^2}\simeq &~\frac{ik_3^3}{4a_B}\frac{\zeta_{k_3}^*(\eta_+)\zeta_{k_2}^{\mathrm{m}\prime}(\eta_{B-})\zeta_{k_3}^{\mathrm{m}\prime}(\eta_{B-})}{\zeta_{k_1}(\eta_+)\zeta_{k_2}(\eta_+)}\left(\frac{\phi_B'}{\phi'(\eta_-)}\right)^4 \left(\frac{B_3(\eta_B)z^2_B}{M_p^2}\frac{\vec{k}_1\cdot\vec{k}_3}{k_3^2}+\frac{B_4(\eta_B)z^4_B}{M_p^4}\frac{(\vec{k}_2\cdot\vec{k}_3)^2}{k_3^2k_2^2}+B_7(\eta_B)\right)\nonumber \\
 &\times\int_{\eta_-}^{\eta_+}d\eta~\left[\zeta_{k_1}^{\mathrm{m}}(\eta_{B-})+\zeta_{k_1}^{\mathrm{m}\prime}(\eta_{B-})\left(\frac{\phi_B'}{\phi'(\eta_{-})}\right)^2(\eta-\eta_{-})\right]~,
\end{align}
where we denote $z_B\equiv z(\eta_B)$.
Performing the integral and using
Eq.\ \eqref{eq:zetabounceNL} for $\zeta_{k_i}(\eta_+)$
(again, this contributes to obtaining an upper bound
for $\mathcal{A}_{\zeta\zeta'^2}$), we obtain
\begin{align}
 \mathcal{A}_{\zeta\zeta'^2}\simeq &~\frac{ik_3^3}{2a_B}\left(\frac{\phi_B'}{\phi'(\eta_-)}\right)^4
 \left(\frac{B_3(\eta_B)z^2_B}{M_p^2}\frac{\vec{k}_1\cdot\vec{k}_3}{k_3^2}+\frac{B_4(\eta_B)z^4_B}{M_p^4}\frac{(\vec{k}_2\cdot\vec{k}_3)^2}{k_3^2k_2^2}+B_7(\eta_B)\right)
 \frac{\zeta_{k_2}^{\mathrm{m}\prime}\zeta_{k_3}^{\mathrm{m}\prime}\zeta_{k_3}^{\mathrm{m}*}}{\zeta_{k_2}^{\mathrm{m}}}
 \Delta\eta_{\mathrm{amp}} \nonumber \\
 &\times\left[1+\frac{\zeta_{k_1}^{\mathrm{m}\prime}}{\zeta_{k_1}^{\mathrm{m}}}\left(\frac{\phi_B'}{\phi'(\eta_-)}\right)^2\Delta\eta_{\mathrm{amp}}\right]
 \left[1+2\frac{\zeta_{k_3}^{\mathrm{m}\prime *}}{\zeta_{k_3}^{\mathrm{m}*}}\left(\frac{\phi_B'}{\phi'(\eta_-)}\right)^2\Delta\eta_{\mathrm{amp}}\right]
 \left[1+2\frac{\zeta_{k_1}^{\mathrm{m}\prime}}{\zeta_{k_1}^{\mathrm{m}}}\left(\frac{\phi_B'}{\phi'(\eta_-)}\right)^2\Delta\eta_{\mathrm{amp}}\right]^{-1} \nonumber \\
 &\times\left[1+2\frac{\zeta_{k_2}^{\mathrm{m}\prime}}{\zeta_{k_2}^{\mathrm{m}}}\left(\frac{\phi_B'}{\phi'(\eta_-)}\right)^2\Delta\eta_{\mathrm{amp}}\right]^{-1}~, \label{eq:Azzp2}
\end{align}
where the modes $\zeta_{k_i}^{\mathrm{m}}$ are implicitly
evaluated at $\eta_{B-}$ and where we recall that
$2\Delta\eta_{\mathrm{amp}}=\eta_+-\eta_-$.
At this point, one could substitute $\zeta_{k_i}^{\mathrm{m}}(\eta_{B-})$
with Eq.\ \eqref{eq:zetam} and write the full expression for
$\mathcal{A}_{\zeta\zeta'^2}$. However, to satisfy the observational
bound on the tensor-to-scalar ratio, we expect there to be a
large amplification of curvature perturbations during the interval
$[\eta_-,\eta_+]$. In fact, from Eqs.\ \eqref{eq:constraintfromr}
and\ \eqref{eq:ubDeltazeta}, it must be that
$|\zeta_{k_i}^{\mathrm{m}\prime}/\zeta_{k_i}^{\mathrm{m}}|[\phi_B'/\phi'(\eta_-)]^2\Delta\eta_{\mathrm{amp}}\gg\mathcal{O}(1)$.
In that limit, the shape function\ \eqref{eq:Azzp2} reduces to (to leading order)
\begin{equation}\label{eq:AIM}
 \mathcal{A}_{\zeta\zeta'^2}\simeq\frac{ik_3^3}{4a_B}\left(\frac{\phi_B'}{\phi'(\eta_-)}\right)^4
 \left(\frac{B_3(\eta_B)z^2_B}{M_p^2}\frac{\vec{k}_1\cdot\vec{k}_3}{k_3^2}+\frac{B_4(\eta_B)z^4_B}{M_p^4}\frac{(\vec{k}_2\cdot\vec{k}_3)^2}{k_3^2k_2^2}+B_7(\eta_B)\right)
 \Delta\eta_{\mathrm{amp}}
 \left|\zeta_{k_3}^{\mathrm{m}\prime}\right|^2~,
\end{equation}
which is purely imaginary, and hence,
does not physically contribute to the physical shape function.
The next-to-leading order terms are
\begin{align}
 \mathcal{A}_{\zeta\zeta'^2}\simeq &~\frac{ik_3^3}{8a_B}\left(\frac{\phi_B'}{\phi'(\eta_-)}\right)^2
 \left(\frac{B_3(\eta_B)z^2_B}{M_p^2}\frac{\vec{k}_1\cdot\vec{k}_3}{k_3^2}+\frac{B_4(\eta_B)z^4_B}{M_p^4}\frac{(\vec{k}_2\cdot\vec{k}_3)^2}{k_3^2k_2^2}+B_7(\eta_B)\right) \nonumber \\
 &\times\frac{\zeta_{k_2}^{\mathrm{m}\prime}\zeta_{k_3}^{\mathrm{m}\prime}\zeta_{k_3}^{\mathrm{m}*}}{\zeta_{k_2}^{\mathrm{m}}}
 \left[\frac{\frac{\zeta_{k_1}^{\mathrm{m}\prime}}{\zeta_{k_1}^{\mathrm{m}}}+2\frac{\zeta_{k_3}^{\mathrm{m}\prime *}}{\zeta_{k_3}^{\mathrm{m}*}}}{\frac{\zeta_{k_1}^{\mathrm{m}\prime}\zeta_{k_2}^{\mathrm{m}\prime}}{\zeta_{k_1}^{\mathrm{m}}\zeta_{k_2}^{\mathrm{m}}}}
 -\frac{\zeta_{k_1}^{\mathrm{m}\prime}\zeta_{k_3}^{\mathrm{m}\prime *}}{\zeta_{k_1}^{\mathrm{m}}\zeta_{k_3}^{\mathrm{m}*}}
 \left(\frac{\zeta_{k_1}^{\mathrm{m}2}\zeta_{k_2}^{\mathrm{m}}}{\zeta_{k_1}^{\mathrm{m}\prime 2}\zeta_{k_2}^{\mathrm{m}\prime}}+\frac{\zeta_{k_1}^{\mathrm{m}}\zeta_{k_2}^{\mathrm{m}2}}{\zeta_{k_1}^{\mathrm{m}\prime}\zeta_{k_2}^{\mathrm{m}\prime 2}}\right)\right]~.
\end{align}
Using Eq.\ \eqref{eq:zetam} for
$\zeta_{k_i}^{\mathrm{m}}(\eta_{B-})$
and taking the limit $k\ll\mathcal{H}$,
we find the leading order real-valued contribution to be
\begin{equation}
\mathcal{A}_{\zeta\zeta'^2}\simeq -\frac{A^2}{16a_B\Delta\eta_{B-}^4}\left(\frac{\phi_B'}{\phi'(\eta_{-})}\right)^2
  (-k_1^3+k_2^3+k_3^3)\left(\frac{B_3(\eta_B)z_B^2}{M_p^2}\frac{\vec{k}_1\cdot\vec{k}_3}{k_3^2}+\frac{B_4(\eta_B)z_B^4}{M_p^4}\frac{(\vec{k}_2\cdot\vec{k}_3)^2}{k_3^2k_2^2}+B_7(\eta_B)\right)~.
\end{equation}
The $B_n(\eta_B)$ terms can be evaluated using Eqs.\ \eqref{eq:B_3_appendix},
\eqref{z2_close_bounce}, and\ \eqref{eq:dotphiB}:
\begin{align}
\label{eq:B3sf}
 \frac{B_3(\eta_B) z_B^2}{M_p^2}\simeq&-\frac{6 \beta  M_p^4  a_B^5 \left(3 \beta  M_p^2 a_B^2+2 \gamma ^2 \phi'^2 _B\right)}{\gamma ^4 \phi'^4 _B}~, \\
\label{eq:B4sf}
 \frac{B_4(\eta_B) z_B^4}{M_p^4}\simeq&-\frac{9 \beta^2 M_p^6 a_B^7 \left(4 M_p^4 \Upsilon a_B^6-3 \gamma ^2 \phi'^{6} _B\right)}{2 \gamma ^6 \phi'^{10} _B}~, \\
\label{eq:B7sf}
 B_7(\eta_B)\simeq&~\frac{3 M_p^2 a^3_B}{2 \gamma^6 \phi'^{10}_B}
  \left(-9a_B^4\beta^2\gamma^2M_p^4\phi_B'^6+6a_B^2\beta\gamma^4M_p^2\phi_B'^8+6\gamma^6\phi_B'^{10}+12a_B^{10}\beta^2M_p^8\Upsilon
    +24a_B^8\beta\gamma^2M_p^6\phi_B'^2\Upsilon\right. \nonumber \\
    &\left.+32a_B^6\gamma^4M_p^4\phi_B'^4\Upsilon\right)~.
\end{align}

Similarly, we can use the previous procedure
to find the contribution from the $\zeta'^3$
term to the shape function (again, omitting
the additional perturbations for now),
\begin{align}\label{eq:Azp3}
\mathcal{A}_{\zeta'^3}=&~\frac{ik_3^3\zeta_{k_3}^*(\eta_+)}{4\zeta_{k_1}(\eta_{+})\zeta_{k_2}(\eta_{+})}\int_{\eta_-}^{\eta_+}d\eta~
  \left(\frac{B_6(\eta)}{a_B^2}\zeta'_{k_1}(\eta)\zeta'_{k_2}(\eta)\zeta'_{k_3}(\eta)\right) \nonumber \\
  =&~\frac{ik_3^3\zeta_{k_3}^*(\eta_+)}{2\zeta_{k_1}(\eta_{+})\zeta_{k_2}(\eta_{+})}
  \frac{B_6(\eta_B)}{a_B^2}\zeta_{k_1}^{\mathrm{m}\prime}\zeta_{k_2}^{\mathrm{m}\prime}\zeta_{k_3}^{\mathrm{m}\prime}
  \left(\frac{\phi'_B}{\phi'(\eta_-)}\right)^6\Delta\eta_{\mathrm{amp}} \nonumber \\
  =&~\frac{ik_3^3B_6(\eta_B)}{2a_B^2}
  \left(\frac{\phi'_B}{\phi'(\eta_-)}\right)^4
  \left\{|\zeta_{k_3}^{\mathrm{m}\prime}|^2
  +\frac{\zeta_{k_3}^{\mathrm{m}\prime}\zeta_{k_3}^{\mathrm{m}*}}{2\Delta\eta_{\mathrm{amp}}}
  \left(\frac{\phi'(\eta_-)}{\phi'_B}\right)^2
  \left[1-\frac{\zeta_{k_3}^{\mathrm{m}\prime *}}{\zeta_{k_3}^{\mathrm{m} *}}
  \left(\frac{\zeta_{k_1}^{\mathrm{m}}}{\zeta_{k_1}^{\mathrm{m}\prime}}+\frac{\zeta_{k_2}^{\mathrm{m}}}{\zeta_{k_2}^{\mathrm{m}\prime}}\right)\right]+...\right\} \nonumber \\
  \simeq&-\frac{A^2B_6(\eta_B)}{16a_B^2\Delta\eta_{\mathrm{amp}}\Delta\eta_{B-}^4}
  \left(\frac{\phi'_B}{\phi'(\eta_-)}\right)^2(k_1^3+k_2^3+k_3^3)~.
\end{align}
The ellipsis in the third line denotes higher-order terms in
$|\zeta_{k_i}^{\mathrm{m}}/\zeta_{k_i}^{\mathrm{m}\prime}|[\phi'(\eta_-)/\phi_B']^2(\Delta\eta_{\mathrm{amp}})^{-1}$,
and in the fourth line, we took the leading order real-valued term
in the expansion.
From Eq.\ \eqref{eq:B_3_appendix}, the $B_6$ term is given by
\begin{equation}
\label{eq:B6sf}
  B_6(\eta_B)\simeq\frac{10 M_p^4 a_B^6 \left[2 \beta M_p^2 a_B^2 + 3 \gamma^2 \phi'^{2}_B \right]}{\gamma^3 \phi'^{5}_B}~.
\end{equation}
From the same argument as for
$\mathcal{A}_{\zeta\zeta'^2}$, the result
\eqref{eq:Azp3} is actually an upper bound
(in absolute value) for $\mathcal{A}_{\zeta'^3}$,
which we will comment on later.

The contribution from the field redefinition
term to the shape function is (again, omitting
the additional perturbations for now)
\begin{align}
\mathcal{A}_{\mathrm{redef}}=&~k_3^3\left[-\frac{A_{18}(\eta_{+}) a_B z^2(\eta_{+})}{4 M_p^4}\left(\frac{\vec{k}_1\cdot(\vec{k}_3-\vec{k}_1)}{k_1^2}-\frac{(\vec{k}_1\cdot\vec{k}_3)[(\vec{k}_3-\vec{k}_1)\cdot\vec{k}_3]}{k_1^2 k_3^2}\right)+\frac{2 A_4(\eta_{+}) a_B^3 - C_1(\eta_{+})}{8 z^2(\eta_{+}) c_s^2}\right]\frac{\zeta'_{k_1}(\eta_+)}{\zeta_{k_1}(\eta_+)} \nonumber \\
  =&~k_3^3\left[-\frac{A_{18}(\eta_{+}) a_B z^2(\eta_{+})}{4 M_p^4}\left(\frac{\vec{k}_1\cdot(\vec{k}_3-\vec{k}_1)}{k_1^2}-\frac{(\vec{k}_1\cdot\vec{k}_3)[(\vec{k}_3-\vec{k}_1)\cdot\vec{k}_3]}{k_1^2 k_3^2}\right)+\frac{2 A_4(\eta_{+}) a_B^3 - C_1(\eta_{+})}{8 z^2(\eta_{+}) c_s^2}\right] \nonumber \\
  &\times\frac{\zeta_{k_1}^{\mathrm{m}\prime}}{\zeta_{k_1}^{\mathrm{m}}}
  \left(\frac{\phi'_B}{\phi'(\eta_-)}\right)^2
  \left[1+\frac{\zeta_{k_1}^{\mathrm{m}\prime}}{\zeta_{k_1}^{\mathrm{m}}}
  \left(\frac{\phi'_B}{\phi'(\eta_-)}\right)^2
  2\Delta\eta_{\mathrm{amp}}\right]^{-1} \nonumber \\
  \simeq&~\frac{k_3^3}{2\Delta\eta_{\mathrm{amp}}}\left[-\frac{A_{18}(\eta_{+}) a_B z^2(\eta_{+})}{4 M_p^4}\left(\frac{\vec{k}_1\cdot(\vec{k}_3-\vec{k}_1)}{k_1^2}-\frac{(\vec{k}_1\cdot\vec{k}_3)[(\vec{k}_3-\vec{k}_1)\cdot\vec{k}_3]}{k_1^2 k_3^2}\right)+\frac{2 A_4(\eta_{+}) a_B^3 - C_1(\eta_{+})}{8 z^2(\eta_{+}) c_s^2}\right]~, \nonumber \\
\end{align}
where we took the leading order term in
$|\zeta_{k_1}^{\mathrm{m}}/\zeta_{k_1}^{\mathrm{m}\prime}|[\phi'(\eta_-)/\phi_B']^2(\Delta\eta_{\mathrm{amp}})^{-1}$.
Here,
\begin{align}
\frac{A_{18}(\eta_+)a_Bz^2(\eta_+)}{M_p^4}&\simeq\frac{3a_B^4\beta M_p^4}{\gamma^3\phi'^{5}(\eta_{+})}~, \\
\frac{2A_4(\eta_+)a_B^3-C_1(\eta_+)}{z^2(\eta_+)c_s^2}&\simeq\frac{4}{\beta\gamma\phi'^3(\eta_+)}\left[3a_B^2M_p^2\beta+2\gamma^2\phi'^2(\eta_+)\right]~.
\end{align}
Recalling Eq.\ \eqref{eq:dotphit}
and the fact that
$|\eta_+-\eta_B|=|\eta_--\eta_B|=\Delta\eta_{\mathrm{amp}}$,
we have $\phi'(\eta_+)=\phi'(\eta_-)$,
and so we can rewrite the terms above as
\begin{align}\label{eq:A18sf}
\frac{A_{18}(\eta_+)a_Bz^2(\eta_+)}{M_p^4}&\simeq\frac{3a_B^4\beta M_p^4}{\gamma^3\phi_B'^{5}}\left(\frac{\phi_B'}{\phi'(\eta_-)}\right)^5~, \\
\label{eq:A4C1sf}
\frac{2A_4(\eta_+)a_B^3-C_1(\eta_+)}{z^2(\eta_+)c_s^2}&\simeq\frac{12a_B^2M_p^2}{\gamma\phi_B'^3}\left(\frac{\phi_B'}{\phi'(\eta_-)}\right)^3
  +\frac{8\gamma}{\beta\phi_B'}\left(\frac{\phi_B'}{\phi'(\eta_-)}\right)~.
\end{align}

Combining the different contributions
(including all permutations), the general form of
the total shape function is found to be
\begin{align}\label{shapefunctionfull}
 \mathcal{A}(k_1,k_2,k_3)\simeq &~
 \frac{-A^2}{16 \Delta\eta_{B-}^4}\left(\frac{\phi'_B}{\phi'(\eta_-)}\right)^2 \frac{1}{\prod_i k_i^2}
 \left[\frac{B_3(\eta_B) z_B^2}{2 M_p^2 a_B}
 \left(2 \sum_{i\neq j} k_i^7 k_j^2 - 2 \sum_{i\neq j} k_i^5 k_j^4 - \sum_{i\neq j\neq \ell} k_i^5 k_j^2 k_\ell^2\right)\right. \nonumber \\
 &+\left.\frac{B_4(\eta_B) z_B^4}{4 M_p^4 a_B}
 \left(-\sum_{i} k_i^9 + 2 \sum_{i\neq j} k_i^6 k_j^3 + 6 \sum_{i\neq j} k_i^7 k_j^2 - 6 \sum_{i\neq j} k_i^5 k_j^4 - 2 \sum_{i\neq j\neq \ell} k_i^4 k_j^3 k_\ell^2 + 2 \sum_{i\neq j\neq \ell} k_i^5 k_j^2 k_\ell^2\right)\right] \nonumber \\
 &-\left[\frac{A^2}{8 \Delta\eta_{B-}^4}\left(\frac{\phi'_B}{\phi'(\eta_-)}\right)^2 \frac{B_7(\eta_B)}{a_B}
 +\frac{3 A^2}{8 \Delta\eta_{B-}^4 \Delta\eta_{\mathrm{amp}}} \left(\frac{\phi'_B}{\phi'(\eta_-)}\right)^2 \frac{B_6(\eta_B)}{a_B^2}
 -\frac{2 A_4(\eta_+) a_B^3 - C_1(\eta_+)}{8 z^2(\eta_+) c_s^2 \Delta\eta_{\mathrm{amp}}}\right] \sum_i k_i^3 \nonumber \\
 &-\frac{A_{18}(\eta_+) a_B z^2(\eta_+)}{32 M_p^4 \Delta\eta_{\mathrm{amp}}} \frac{1}{\prod_i k_i^2}
 \left(\sum_{i\neq j} k_i^7 k_j^2 - 2\sum_{i\neq j} k_i^5 k_j^4 - 2\sum_{i\neq j\neq \ell} k_i^5 k_j^2 k_\ell^2 + \sum_{i\neq j} k_i^6 k_j^3 - \sum_{i\neq j\neq \ell} k_i^4 k_j^3 k_\ell^2\right)~.
\end{align}

\end{document}